\documentclass[authoryear,preprint,12pt]{elsarticle} 

\usepackage{amssymb,amstext,amsmath}
\usepackage{empheq}
\usepackage{comment}
\usepackage{graphicx}
\usepackage[colorlinks]{hyperref}
\usepackage{stmaryrd} 
\usepackage{booktabs}
\usepackage{array}

\usepackage[utf8]{inputenc}
\usepackage[T1]{fontenc}
\usepackage[a4paper,margin=25mm]{geometry}

\newcommand{\boldface}[1]{\boldsymbol{#1}}  

\newcommand{\bfd}{\boldface{d}}
\newcommand{\bfe}{\boldface{e}}

\newcommand{\bfn}{\boldface{n}}

\newcommand{\bfp}{\boldface{p}}
\newcommand{\bfq}{\boldface{q}}

\newcommand{\bft}{\boldface{t}}
\newcommand{\bfu}{\boldface{u}}
\newcommand{\bfv}{\boldface{v}}
\newcommand{\bfw}{\boldface{w}}
\newcommand{\bfx}{\boldface{x}}

\newcommand{\bfG}{\boldface{G}}

\newcommand{\bfI}{\boldface{I}}

\newcommand{\bfR}{\boldface{R}}

\newcommand{\bfV}{\boldface{V}}

\newcommand{\bfY}{\boldface{Y}}

\newcommand{\bfepsvar}{\boldsymbol{\epsilon}}

\newcommand{\bfepsilon}{\boldsymbol{\varepsilon}}

\newcommand{\bfxi}{\boldsymbol{\xi}}
\newcommand{\bfsigma}{\boldsymbol{\sigma}}

\newcommand{\bfvarphi}{\boldsymbol{\varphi}}

\newcommand{\bfchi}{\boldsymbol{\chi}}

\newcommand{\bfnull}{\boldsymbol{0}}

\newcommand{\bfPsi}{\boldsymbol{\Psi}}

\newcommand{\calE}{\mathcal{E}}
\newcommand{\calF}{\mathcal{F}}
\newcommand{\calG}{\mathcal{G}}

\newcommand{\calP}{\mathcal{P}}

\newcommand{\calR}{\mathcal{R}}
\newcommand{\calS}{\mathcal{S}}

\newcommand{\dsR}{\mathbb{R}}

\newcommand{\dsZ}{\mathbb{Z}}

\newcommand{\sfbC}{\boldsymbol{\mathsf{C}}}
\newcommand{\sfbD}{\boldsymbol{\mathsf{D}}}

\newcommand{\sfbR}{\boldsymbol{\mathsf{R}}}
\newcommand{\sfbS}{\boldsymbol{\mathsf{S}}}

\newcommand{\sfD}{\mathsf{D}}

\newcommand{\sfI}{\mathsf{I}}

\newcommand{\sfR}{\mathsf{R}}
\newcommand{\sfS}{\mathsf{S}}

\newcommand{\totderiv}[2]{\frac{\dd #1}{\dd #2}}

\newcommand{\half}{\frac{1}{2}}
\newcommand{\T}{^{\mathrm{T}}} 

\newcommand{\me}{^{\mathrm{-1}}} 
\newcommand{\Rset}{\mathbb{R}}

\newlength{\boxwidth}
\newcommand\pl{\partial}
\def\dd{\;\!\mathrm{d}}

\def\btheorem{\begin{theorem}}
\def\etheorem{\end{theorem}}
\def\blemma{\begin{lemma}}
\def\elemma{\end{lemma}}
\def\bproposition{\begin{proposition}}
\def\eproposition{\end{proposition}}
\def\bcorollary{\begin{corollary}}
\def\ecorollary{\end{corollary}}
\def\bdefinition{\begin{definition}}
\def\edefinition{\end{definition}}
\def\bexample{\begin{example}}
\def\eexample{\end{example}}
\def\bremark{\begin{remark}}
\def\eremark{\end{remark}}

\def\ol{\overline}

\def\ds{\displaystyle}

\DeclareMathOperator{\sign}{sign}

\newcommand{\be}{\begin{equation*}}
\newcommand{\ee}{\end{equation*}}

\newcommand{\beq}{\begin{eqnarray*}}
\newcommand{\eeq}{\end{eqnarray*}}
\newcommand{\bem}{\begin{multline}}
\newcommand{\eem}{\end{multline}}
\newcommand{\ba}{\begin{align*}}
\newcommand{\ea}{\end{align*}}

\newcommand{\fp}[2]{\frac{\partial #1}{\partial #2}}

\usepackage{IEEEtrantools} 
\usepackage{cases}

\newcommand{\fwdFT}[1]{\calF\{#1\}}
\newcommand{\invFT}[1]{\calF\me\left\{#1\right\}}

\newcommand{\nstp}{^{(i)}}
\newcommand{\bfEpsS}{\bfepsvar^\mathrm{s}}
\newcommand{\EpsS}{\epsilon^\mathrm{s}}

\usepackage{amsthm}
\newtheorem*{remark}{Remark}
\newcommand{\parT}{\partial_t}
\newcommand{\ab}{{\alpha \beta}}
\newcommand{\beal}{{\beta \alpha}}
\newcommand{\sect}{Section}

\newcommand{\degree}{^\circ}
\hypersetup{
	citecolor=blue,
	linkcolor=blue,
	urlcolor=blue
}

\journal{Computer Methods in Applied Mechanics and Engineering}

\begin{document}

\begin{frontmatter}



\title{An electromechanically coupled multiphase-field model with generalized kinetic relations for ferroelectrics}


\author[inst1]{Hsu-Cheng Cheng}

\affiliation[inst1]{organization={Mechanics \& Materials Laboratory},
            addressline={Department of Mechanical and Process Engineering}, 
            state={ETH~Z{\"u}rich},
            postcode={8092},
            state={Z{\"u}rich},
            country={Switzerland}}

\author[inst1]{Dennis M. Kochmann \corref{cor1}}

\cortext[cor1]{Corresponding author}

\begin{abstract}
Classical phase-field models typically use the Allen--Cahn evolution law to model interface motion as a consequence of energy-gradient descent. This implies a linear kinetic relation between the velocity of an interface (e.g., a domain wall in a ferroelectric) and its driving force. When nonlinear kinetic relations are to be modeled, as commonly found in ferroelectric ceramics, an alternative evolution law and hence an alternative model is needed. Here, we propose an electromechanically (stress-driven) coupled multiphase-field framework with a general kinetic formulation, which integrates a prescribed kinetic relation directly into the evolution law. We show that this model correctly evolves domain walls with the assigned nonlinear kinetics, e.g., of mixed exponential-power law type, following the so-called Merz--Stadler law. The multiphase formulation further enables distinct kinetic relations and interfacial energies to be assigned to different order-parameter pairs, which reflects the distinguishable properties of the different types of ferroelectric domain walls. The proposed framework is broadly applicable for simulating kinetic relations in materials with applications beyond ferroelectrics.
\end{abstract}




\end{frontmatter}


\section{Introduction}\label{sec:intro}

Ferroelectric materials have been exploited for their prominent electromechanically coupled responses, which enable innovations in, among others, dielectric energy storage \citep{yang2019perovskite}, biological applications \citep{blazquez2018biological}, tunable resistors \citep{chanthbouala2012ferroelectric}, and memory devices \citep{uchino2018ferroelectric, troiler2020impact}. Their outstanding electromechanical response is based on the formation of spontaneous polarization and its ability to switch under applied fields \citep{lines2001principles}. Perovskite ceramics are among the most technologically relevant ferroelectrics. Below their Curie temperature, the spontaneous polarization (described by the polarization vector $\bfp^s$) is formed due to atomic interactions at the unit cell level \citep{rabe2007physics} and causes an associated spontaneous strain (described by the strain tensor $\bfepsilon^s$, measured relative to the unit cell's unpolar cubic configuration). Owing to crystallographic symmetries, multiple preferred unit cell configurations and hence spontaneous polarizations compete at room temperature. In a tetragonal perovskite, e.g., six unique polarizations exist in the polar state. A \textit{domain} refers to a region of uniform spontaneous polarization vector. Domains are separated by \textit{domain walls}, atomically sharp regions that separate two distinct domains. To minimize the electrostatic and elastic energy, domain walls tend to orient such that both the normal component of spontaneous polarization and the strain discontinuity across the wall vanish. These compatibility conditions together constrain the allowed wall orientation to specific crystallographic planes. A $180\degree$ wall separates domains with antiparallel spontaneous polarization vectors and is commonly termed a \textit{ferroelectric} domain wall, since the spontaneous strain is continuous across the wall. For tetragonal perovskites, a $90\degree$ domain wall that orients at a 45$\degree$ to the spontaneous polarization possesses compatible strains and is commonly termed a \textit{ferroelectric/ferroelastic} domain wall. Any other orientation constitutes a \textit{charged domain wall}, which further exhibits mechanical incompatibility for 90$\degree$ domain walls. The dielectric and piezoelectric functionality of a ferroelectric material is based on the (re-)distribution of domains under applied mechanical or electrical loads. Domain wall motion is one of the key mechanisms in this process, which is why domain walls and their evolution under external loading have been widely investigated \citep[see, e.g.,][]{tagantsev2010,meier2020domain,schultheiss2023ferroelectric}. 

Domain-wall kinetics have been reported for single-crystal barium titanate in the pioneering works of \citet{miller1958velocity, miller1959, taylor1962high, Stadler1963, stadler1964temperature, stadler1966forward}, and summarized by \citet{schultheiss2023ferroelectric}. Studying a single crystal isolates the intrinsic domain-wall response from grain interactions in a polycrystal, which can considerably complicate the relation between macroscopic performance and domain-level microstructure evolution. Those studies revealed that varying the amplitude $e$ of an applied electric field can directly alter the velocity $V$ of a $180\degree$ domain wall during its sideways motion. In particular, two distinct regimes were reported, separated by a critical field amplitude of about $e \approx 100~\mathrm{kV}/\mathrm{m}$. Below this value, the velocity follows an exponential dependence, $V \propto \exp\!\left(-e_a/e\right)$, where $e_a$ is the activation field. This relation is known as \textit{Merz' law}. Above $e_a$, the dependence follows a power law $V \propto e^{r}$, whose constant $r$ was found to be around 1.4. This relation is also known as \textit{Stadler's law}. These early-stage experimental indication of a nonlinear kinetic relation of ferroelectric domain walls was later confirmed also by molecular dynamics simulations \citep{liu2016intrinsic, boddu2017molecular} and implied by the rate-dependent studies \citep{schultheiss2018revealing, kannan2022kinetics}

Beyond electric loading, mechanical stress has been widely recognized to also affect the domain evolution and resulting polarization switching in ferroelectrics \citep{kumazawa1998effect,gruverman2003mechanical, burcsu2004large}. Moreover, the local (electrical and mechanical) fields near a domain wall may differ significantly from the macroscopically applied fields, so that even in purely electrical loading, the local driving force may involve mechanical fields. Overall, this calls for a kinetic relation that, besides the electric field, also depends on the mechanical stress (or strain) state. One possibility is to adopt a mechanistic description that relates the domain wall motion to a thermodynamic driving force \citep{jiang1994driving, abeyaratne2006evolution}, defined as the configurational force derived from the second law of thermodynamics. Under the assumption of isotropic electrostatics (while neglecting piezoelectric coupling and all mechanical fields), Merz' law for $180\degree$ domain walls can be rewritten as the kinetic relation of this thermodynamic driving force \citep{Guin2023phase}. However, this neglects mechanics and especially the piezoelectric coupling, which introduces additional stress effects in the driving force. As a result, models that account for the generally nonlinear, electromechanically coupled kinetic relation of domain wall motion have remained an open challenge. 

Phase-field methods are effective tools for simulating interface motion and its relation to microstructure. Their application to ferroelectric materials has enabled studies of correlations between domain evolution and material properties \citep{zhang_Bhattacharya_2005,su_landis_2007,chen2008phase}, including domain interactions with grain boundaries \citep{choudhury2007effect,su2015phase,indegand2023Grain} and other defects \citep{zuo2014domain, shindo2015phase, vorotiahin2020hierarchy, indergand2021effect}. Additional studies were reviewed comprehensively by \citet{wang2019understanding, schultheiss2023ferroelectric}. As discussed by \citet{Guin2023phase}, current phase-field models for ferroelectric domain wall motion typically employ an Allen-Cahn-type evolution law (a gradient descent that aims to minimize the free energy or electric enthalpy). For low to moderate applied field magnitudes, this evolution law implies a \textit{linear} kinetic relation, which cannot capture the experimentally observed nonlinear kinetics of domain wall motion. Therefore, a phase-field model that allows for a given (generally nonlinear) kinetic relation in ferroelectrics is needed but presently missing. Note that such a phase-field model holds potential beyond the specific material systems discussed here, since nonlinear kinetic relations may govern phase kinetics in various other applications such as martensitic phase transformations \citep{abeyaratne1997kinetics,abeyaratne2006evolution}.

Several studies have incorporated nonlinear kinetic relations into phase-field models. \citet{liang2012nonlinear} modified the Allen-Cahn equation by introducing an exponential dependence on the variational derivative for electrode-electrolyte interface evolution. Threshold-type kinetic responses have also been achieved through variational formulations based on dissipation potentials. \citet{tuuma2018rate} employed rate-independent dissipation potentials to model shape-memory phase transformations, while \citet{gokuli2021_Brandon} used a minimum-dissipation-potential framework to model grain boundary migration. These models are restricted to threshold-type kinetic relations. In ferroelectrics, we require a phase-field model that evolves the domain walls with a mixed exponential-power relation and possibly other kinetic relations that remain to be identified. We therefore seek a phase-field evolution law that can incorporate any (physically sensible) given kinetic relation.  

The level-set method is an alternative to phase-field models, which simulates interface motion via implicit interface tracking \citep{osher1988, sethian1996fast, osher_level_2003}. The interface is represented as the zeroth level set of a signed-distance function $\psi$. Its evolution is governed by the kinematic equation $\parT \psi + V\lvert\nabla\psi\rvert = 0$ \citep{osher1988}, where $V$ denotes the normal velocity of the level set. Since this evolution law is written directly in terms of the scalar quantity $V$, the level-set formulation admits prescribing the desired kinetic relation by linking $V$ to a configurational driving force. Such driving forces were shown to follow the second law of thermodynamics in close connection with sharp-interface theory, with applications including martensitic transformations \citep{hou1999level_Rosakis} and ferroelectrics \citep{cheng2025fft}. However, using the level-set method requires frequent, precise reinitialization of the signed-distance function to ensure accurate kinetics. When multiphase evolution is to be simulated, such as in ferroelectrics, the extension to multiphase evolution with nonlinear kinetics is nontrivial \citep{zhao1996variational_Osher, SAYE_Sethian}. 

Explicitly incorporating kinetic relations in the kinematics of level sets has motivated phase-field models whose evolution laws replace the signed-distance function with that of an order parameter. A key distinction between the level-set method and a phase-field model lies in the implicit function used to represent the interface. Phase-field model embeds a diffuse interface directly in the energy formulation, whereas the level-set method identifies the interface as the zeroth level set of the signed-distance function. Consequently, the phase-field model evolves interfaces of approximately constant thickness, while the level-set method propagates all level sets with their associated velocity, followed by recurring reinitialization to reconstruct the signed-distance function. \citet{agrawal2015dynamic} proposed this type of phase-field model for two-phase evolution (with a single order parameter) by using the level-set kinematic equation with the phase-field's order parameter. In their work, the phase-field profile is regulated by introducing a hyperbolic tangent function in the driving force. The extension of this model to multiphase evolution, however, faces challenges analogous to those encountered in the level-set method. Another approach was proposed by \citet{alber2013,alber2016alternative}. In their hybrid model for two-phase evolution, the kinematics of the level-set method is used for the evolution law of a phase-field. The double-well potential is incorporated into the model in a manner consistent with the classical phase-field model to ensure the preservation of the phase-field profile. Asymptotic analysis demonstrated potential benefits over the classical Allen-Cahn model when the interface energy of the problem is low. Despite the model's advantages, it was introduced for two-phase evolution and is not directly applicable to simulate multiphase evolution such as in ferroelectrics. An extension to multiphase evolution was proposed by \citet{Guin2023phase}. Studying traveling-wave solutions for a two-phase interface, they confirmed that the evolution law admits the assignment of a given kinetic relation, and the model is advantageous over the Allen-Cahn model in simulating the evolution of interfaces with low interface energy. 

In this work, inspired by the purely isotropic, electrostatic formulation of \citet{Guin2023phase}, we develop a general kinetic model in the fully electro-mechanically coupled multiphase-field setting, for which we derive the evolution law from thermodynamics and ensure its consistency. In addition, we introduce electromechanical coupling to capture the physics of real-world ferroelectric materials. We further formulate electromechanical coupling within a stress-driven formulation \citep{cheng2025fft} and integrate it into the multiphase-field framework, which allows not only for the direct application of stresses but also for a more efficient Fourier-Galerkin implementation \citep{vondvrejc2014fft} and a unified regularization of the free energy density compared to the strain-driven counterpart. 

The remainder of this paper is structured as follows. \autoref{sec:SI_review} reviews the sharp-interface description of a ferroelectric continuum along with its stress-driven formulation (based on the free energy density) and the thermodynamic driving force on an interface. In \autoref{sec:multiphase}, we derive the evolution law of the multiphase-field model for ferroelectrics, which utilizes the kinematics of level set evolution. The setting enables the derivation of the regularized thermodynamic driving force, which ensures thermodynamic consistency. Numerical studies are presented in \autoref{sec:results}, where different nonlinear kinetic relations are prescribed, and the resulting multiphase evolution is illustrated and validated. Finally, \autoref{sec:summary} concludes our study.

\section{Sharp-interface continuum description of ferroelectrics} \label{sec:SI_review}
In this section, we review the continuum modeling of a ferroelectric solid with a sharp interface that represents a domain wall. This treatment rests on the assumption of a separation of scales, such that the physics inside the domain wall is captured in a phenomenological way by the energetics and kinetics of the sharp interface. The general notion behind this approach relies on the classical balance laws in a continuum \citep{Gurtin_evolvingPhase,abeyaratne2006evolution}, considering an interface as a discontinuity that dissipates energy as it moves. As we focus exclusively on ferroelectrics in this work, the terms ``\textit{domain wall}'' and ``\textit{interface}'' are used interchangeably. 

\subsection{Balance laws in a ferroelectric material}
Let us consider a continuum solid represented as an open-bounded set $\Omega$ in three-dimensional Euclidean space $\dsR^3$ with its boundary $\partial \Omega$. This electromechanically coupled material is subjected to quasistatic loading throughout this study.  The primary variables describing this solid are the displacement field $\bfu(\bfx,t)$ and the electric potential $\phi(\bfx, t)$, both of which are continuous in $\bfx\in\Omega$ (almost everywhere) and time $t$. The deformation of the solid is sufficiently described by the infinitesimal strain tensor%
\footnote{We use the Einstein summation rule and adopt the following notation for tensor calculations: for any vector fields $\bfv(\bfx,t)$ and $\bfw(\bfx,t)$, second-order tensor fields $\bfG(\bfx,t)$ and $\bfY(\bfx,t)$, third-order tensor $\sfbD(\bfx,t)$ and fourth-order tensor $\sfbS(\bfx,t)$, we denote in a Cartesian basis
\begin{equation*}
\begin{aligned}
&\bfv \cdot \bfw=v_i w_i, ~(\nabla \bfv)_{ij}=v_{i,j}, ~\nabla \cdot \bfv=v_{i,i}, ~(\nabla \times \bfv)_i= \calE_{ijk} v_{j,k}, ~(\nabla \cdot \bfG)_i=G_{ij,j}, ~(\bfR \bfv)_i= R_{ij}v_j, \\ 
&(\bfG \bfY)_{ij}=G_{ik} Y_{kj},~\bfG \cdot \bfY=G_{ij} Y_{ij}, ~(\sfbD\T \bfv)_{ij}= \sfD_{kij} v_k, ~(\sfbD \bfG)_{i}= \sfD_{ijk} G_{jk}, ~(\sfbS  \bfG)_{ij}= \sfS_{ijkl} G_{kl}.
\end{aligned}
\end{equation*}}
$\bfepsilon = \tfrac{1}{2}(\nabla\bfu +\nabla\bfu\T) = \nabla^s \bfu$ due to the brittle ceramic nature of the material. Meanwhile, it is assumed that the material is a perfect dielectric, so there are no free charges in $\Omega$. This indicates that the electric field $\bfe(\bfx,t)$ is conservative and can be represented as $\bfe = -\nabla \phi$.

In the absence of any loading, the material in equilibrium exhibits spontaneous polarization states $\bfp^s(\bfx,t)$, originating directly from minimum-energy states of the atomic unit cell. The total polarization $\bfp(\bfx, t)$ follows from the additive decomposition $\bfp = \epsilon_0\bfchi \bfe + \bfp^s$, where $\bfchi(\bfx,t)$ is the second-order electric susceptibility tensor and $\epsilon_0$ is the permittivity of vacuum%
\footnote{For a dielectric material, $\bfepsvar = \epsilon_0 (\bfI +\bfchi)$ is its permittivity, where $\bfI$ is the second-order identity tensor.}.
A concomitant effect of $\bfp^s$ is the spontaneous strain $\bfepsilon^s$, which leads to the analogous decomposition $\bfepsilon = \bfepsilon^e + \bfepsilon^s$, where $\bfepsilon^e$ is the elastic strain of the material.

At equilibrium, the material must satisfy linear momentum balance and Gauss' law, respectively,
\begin{equation} \label{eq:fieldBalance}
    \nabla \cdot \bfsigma = 0, \qquad \nabla \cdot \bfd = 0,
\end{equation}
where $\bfsigma$ is the symmetric Cauchy stress tensor and $\bfd$ the electric displacement vector. 

\begin{figure}[!b]
    \centering
    \includegraphics[width=1\linewidth]{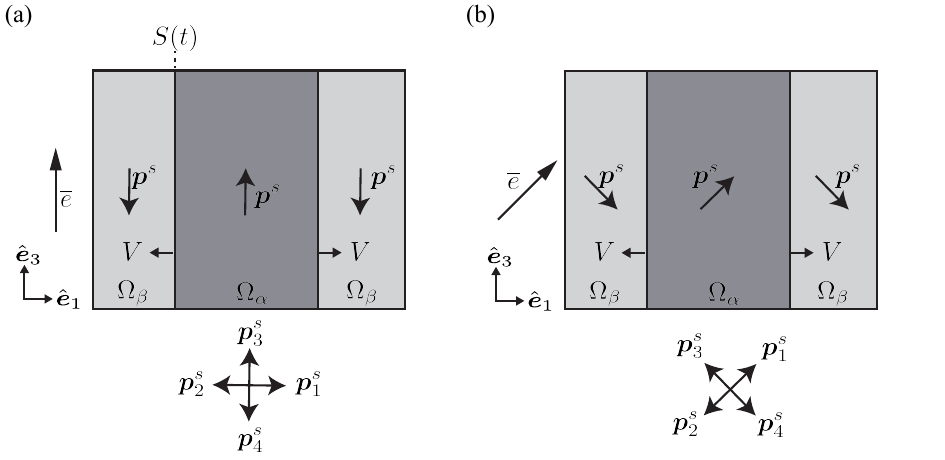}
    \caption{Schematics of (a) 180$\degree$ and (b) 90$\degree$ charge-neutral, straight domain walls. Application of the indicated electric field of magnitude $\ol{e}$ leads to the domain walls moving with a normal velocity magnitude $V$, as shown. Indicated underneath both domain wall configurations are the spontaneous polarization vectors of the four in-plane domain variants (one case being rotated with respect to the other by 45$\degree$).
    } 
    \label{fig:dw_180_90}
\end{figure}

Domains and domain walls naturally form in a ferroelectric material as a consequence of minimizing the elastic and electrostatic energy. We denote a domain by $\Omega_\alpha(t) \subseteq \Omega$ and a domain wall (interface) by $S(t)$, as illustrated in \autoref{fig:dw_180_90}. Although field quantities may experience discontinuities across $S(t)$, the balance laws and compatibility of the material enforce the jump conditions
\begin{equation}
    \llbracket \bfsigma \rrbracket \bfn = 0,  \qquad  \llbracket \nabla \bfu \rrbracket \bft = 0,
\end{equation}
where $\llbracket \bfsigma \rrbracket = \bfsigma^+ - \bfsigma^-$ indicates the jump across the interface, with $\bfsigma^\pm$ indicating the stress in the immediate vicinity of the domain wall on either side, $\bfn$ denotes a unit vector normal to the interface pointing from the $-$ to the $+$ side, and $\bft$ is a unit vector tangent to the interface. The absence of a magnetic field and of free charges further ensures that
\begin{equation}\label{eq:jumpCond_elec}
  \llbracket \bfd \rrbracket \cdot \bfn = 0, \qquad \llbracket \bfe \rrbracket \cdot \bft = 0.
\end{equation}
%
%

\subsection{Stress-driven formulation} \label{sec:StressDriven}
The spontaneous polarization of a domain is constant at a given temperature, and its energy corresponds to one of the minima of a multi-well potential \citep{Devonshire1954, schrade2013physical}. We adopt the approximation of a linear piezoelectric response, when the domain is subjected to electromechanical loading. In particular, we focus on a stress-driven approach \citep{cheng2025fft} to study domain evolution, which aligns with common experimental setups \citep{baither2001ferroelastic,gruverman2003mechanical,kumazawa1998effect, burcsu2004large, wojnar2014broadband, kannan2022kinetics}. In this setting, the Gibbs free energy density $\calG_\alpha$ in a domain $\Omega_\alpha$ is written as 
\begin{equation}\label{eq:gibbs_domain}
    \calG_\alpha (\bfsigma, \bfe) = -\frac{1}{2}\bfe \cdot \bfEpsS_\alpha \bfe - \bfp^s_\alpha \cdot \bfe -\half \bfsigma \cdot \sfbS_\alpha \bfsigma - \bfsigma \cdot \bfepsilon^s_\alpha - \bfsigma \cdot \sfbD_\alpha\T \bfe, 
\end{equation}
where $\bfEpsS_\alpha$ is the second-order dielectric tensor for constant stress, $\sfbS_\alpha$ the fourth-order elastic compliance tensor, and $\sfbD_\alpha$ the third-order strain-charge piezoelectric tensor. The constitutive relations follow as
\begin{equation}
    \label{eq:Constitutive_StressDriv}
    \begin{aligned}
        \bfepsilon &= -\fp{\calG_\alpha}{\bfsigma} = \sfbS_\alpha \bfsigma +\bfepsilon^s_\alpha +\sfbD_\alpha\T \bfe,\\
        \bfd& = -\fp{\calG_\alpha}{\bfe}=\bfEpsS_\alpha \bfe + \bfp^s_\alpha + \sfbD_\alpha \bfsigma.
    \end{aligned}
\end{equation}

The evolution of the domain wall is modeled by recourse to continuum theory of phase transformations, in which the interface motion is linked to the dissipation inside the material via the Clausius-Duhem inequality localized on the domain wall. This inequality takes the form \citep{jiang1994driving, abeyaratne2006evolution}
\begin{equation}\label{eq:CD_sharp}
    \int_S f_s V \dd A \geq 0, 
\end{equation}
where $V$ is the normal velocity of the domain wall (in the direction of $\bfn$), and $f_s$ is the thermodynamically conjugate driving force. For the free energy density in Eq.~\eqref{eq:gibbs_domain}, the driving force in this stress-driven formulation is given by \citep{cheng2025fft} 
\begin{equation}
    \label{eq:SI_DF_stressDriven}
      f_s(\bfsigma, \bfe) = \llbracket \calG \rrbracket +\llbracket \bfsigma \rrbracket \cdot \langle \bfepsilon \rangle + \langle \bfd \rangle \cdot \llbracket \bfe \rrbracket, 
\end{equation}
where $\langle \bfepsilon \rangle=\tfrac{1}{2}(\bfepsilon^+ +\bfepsilon^-)$ denotes the average of the strain tensor across the domain wall. 
Inequality \eqref{eq:CD_sharp} requires a functional relation between $V$ and $f_s$. The physical interpretation of the configurational force motivates formulating a kinetic relation $\check{V}(f_s)$, whose exact form must be obtained from experiments for a particular type of domain wall. This kinetic relation can be considered as an additional constitutive relation to close the problem of domain wall evolution, and as a phenomenological description of the domain evolution encoding a lower-scale physics. In addition, as discussed in \autoref{sec:intro}, this relation is likely nonlinear in ferroelectrics, as, e.g., the Merz-Stadler law confirmed.  


%
\section{Multiphase framework formulation} \label{sec:multiphase}
In this section, we derive a multiphase framework for ferroelectrics that accounts for $M$ distinct domains, which is based on an extension of the two-phase hybrid model of \citet{alber2016alternative}. The derivation combines the multiphase-field concept of \citet{steinbach1999generalized} with level-set kinematics. The resulting evolution law differs from that of \citet{Guin2023phase}, which may be viewed as a simplified version of the present formulation. In phase-field models, sharp interfaces (i.e., domain walls in our context) are approximated by a regularized, smooth, and fast transition over a small but finite interface region \citep{Plapp2012, steinbach2023lectures}. Hence, we interpret the regularized domain wall (having a thickness greater than the real one) as a purely numerical construct introduced to implicitly track the domain wall position.

\subsection{The evolution law} \label{sec:multiphase_evo}
We introduce the order parameter $\varphi_\alpha(\bfx, t) \in [0,1]$ as the volume fraction of a specific domain variant with polarization $\bfp^s_\alpha$ at position $\bfx$ in the ferroelectric at time $t$. We further define the list of fields $\bfvarphi = \{\varphi_1, \varphi_2, \dots, \varphi_M\}$, which is subject to the constraint
\begin{equation}\label{eq:constraint_phi}
    \sum_{\alpha=1}^M\varphi_\alpha(\bfx,t)=1.
\end{equation}
Using these order parameters, we compose a spontaneous polarization $\bfp^s$ at a given location and time as 
\begin{equation} \label{eq:pol_construct}
    \bfp^s(\bfx, t) = \sum_{\alpha=1}^M\varphi_\alpha(\bfx,t) \bfp^s_\alpha.
\end{equation}
The constraint \eqref{eq:constraint_phi} implies a dependence among the order parameters. Consider, e.g., a two-phase interface, separating domains $\alpha$ and $\beta$, which moves over time; the two order parameters $\varphi_\alpha$ and $\varphi_\beta=1-\varphi_\alpha$ evolve dependently, increasing one volume fraction at the expense of the other. Similarly, \eqref{eq:constraint_phi} imposes a dependence of three or more volume fractions near junctions, at which multiple phases intersect.

This dependence motivates a reformulation in terms of \textit{interface fields} \citep{steinbach1999generalized}. To this end,  we adopt the list of fields $\bfPsi$, whose elements $\Psi_{\alpha\beta}$ ($\alpha,\beta=1,\ldots,M$) are linked to the multiphase-field via the linear transformation 
\begin{equation} \label{eq:interfaceField}
    \Psi_{\alpha \beta} =  \varphi_\alpha -\varphi_\beta,
\end{equation}
which is antisymmetric: $\Psi_{\beta \alpha}= - \Psi_{\alpha \beta}$. The inverse transformation is obtained by enforcing the constraint \eqref{eq:constraint_phi} and is given by 
\begin{equation} \label{eq:inverseTrans_phi}
     \varphi_\alpha = \frac{1}{m} \left(\sum_{\beta = 1}^{m} \Psi_{\alpha \beta}+1 \right),
\end{equation}
where $m$ indicates the number of phases involved in the regularized interface region, e.g., $m=2$ at the typical interface region and $m\geq3$ at the junctions. As shown by \citet{steinbach1999generalized}, this linear transformation eliminates the need for Lagrange multipliers in the gradient descent evolution law, which are otherwise required to enforce the constraint \citep{garcke1998anisotropic}. Using Eq.~\eqref{eq:inverseTrans_phi}, the evolution of the order parameter becomes 
\begin{equation}\label{eq:PartialVarphi}
     \parT \varphi_\alpha = \frac{1}{m} \sum_{\beta = 1}^{m} \parT\Psi_{\alpha \beta},
\end{equation}
where $\parT$ denotes the partial time derivative.

One can think of the evolution of a given interface field $\Psi_{\alpha \beta}$ as the transport of the level sets of this field \citep{alber2013,agrawal2015dynamic, alber2016alternative}. This is akin to the level-set method \citep{osher1988, sethian1996fast,osher_level_2003}, in which the level sets of a signed distance function are transported with the normal velocity defined on the given level set\footnote{The equation of motion for a level-set function $\psi$ is classically derived as follows \cite{osher1988}. The level set value of a material point on the front with path $\bfx(t)$ must be constant for all times, i.e., $\psi(\bfx(t), t) = c$ with a constant $c\in\Rset$. Assume that $V=\totderiv{x}{t} \cdot \bfn$ is the known outward normal velocity of the moving level set, where $\bfn = \nabla\psi/\lvert \nabla\psi\rvert$ is the outward unit normal vector. The total time derivative of $\psi$ yields, after insertion of the above relations and algebraic manipulation, the level set equation $\parT\psi + V\lvert\nabla\psi\rvert= 0$.  
}. This motivates us to define the evolution of $\Psi_{\alpha \beta}$, analogous to the traditional level set equation, as 
\begin{equation}\label{eq:evo_Psi}
     \parT\Psi_{\alpha \beta} = V_{\beta \alpha} \lvert\nabla \Psi_{\alpha \beta} \rvert =  V_{\beta \alpha} \sqrt{\left(\nabla\varphi_\alpha - \nabla \varphi_\beta\right) \cdot \left(\nabla\varphi_\alpha - \nabla \varphi_\beta\right) }, 
\end{equation}
where $|\cdot|$ denotes the Euclidean norm and $V_{\beta \alpha} \in \mathbb{R}$ is the normal velocity of the level sets $\Psi_{\alpha \beta}$. Note that normal vector $V_{\beta \alpha}$ points 
from $\Omega_\beta$ to $\Omega_\alpha$ due to the definition~\eqref{eq:interfaceField}, and $V_{\beta \alpha} = - V_{ \alpha \beta}$. Insertion into \eqref{eq:PartialVarphi} leads to the evolution of the order parameter as
\begin{equation}\label{eq:evo_phi_pre}
    \parT\varphi_\alpha = \frac{1}{m} \left(\sum_{\beta = 1}^{m} V_{\beta \alpha}  \sqrt{\left(\nabla\varphi_\alpha - \nabla \varphi_\beta\right) \cdot \left(\nabla\varphi_\alpha - \nabla \varphi_\beta\right) } \right).
\end{equation}
Note that, for a two-phase interface with $m=2$ and $\varphi_\alpha=1-\varphi_\beta$, the above reduces to the classical level set equation $\parT\varphi_\alpha = V_{\beal} \lvert\nabla\varphi_\alpha\rvert$ with level set normal velocity $V_{\beta \alpha}$.

Eq.~\eqref{eq:evo_phi_pre} represents the net evolution of $\varphi_\alpha$, obtained by summing the contributions from each interface pair $\{\varphi_\alpha, \varphi_\beta\}$. In other words, the evolution of an order parameter is decomposed into pairwise interactions with other order parameters in the interface region. The normal velocity $V_{\beal}$, which we assume known here, is generally driven by a thermodynamic driving force $f_{\beal}$, which motivates defining a kinetic relation $V_{\beal}=\check{V}_\beal(f_\beal)$. This kinetic relation, which is explicitly integrated into the evolution equation, encodes the underlying thermodynamic process of the material, as we will show in the following section.   

\subsection{The thermodynamically informed, regularized driving force}
To condense the notation and highlight the phase-field construction derived in this section, we restrict the bulk physics to its mechanical contribution. The electro-mechanical coupling in the bulk can be incorporated into the model via the free energy density in \eqref{eq:reg_drivingforce}, as shown in \citep{cheng2025fft}. Therefore, without loss of generality, the purely mechanical result carries directly over to the full electromechanically coupled material. 

In the classical phase-field formulation, the order parameter smoothly interpolates between the energy of each phase participating in an interface. Therefore, the regularized total energy includes a gradient contribution that penalizes spatial variations of the order parameter, leading to an interfacial phase profile that propagates through the domain in a standing-wave-like manner \citep{Plapp2012, steinbach2023lectures}. 

Following the discussion in \sect~\ref{sec:multiphase_evo}, the total regularized free energy density can be represented as  $\calG\left(\bfsigma,\bfPsi\left(\bfvarphi\right),\nabla\bfPsi \left(\nabla \bfvarphi\right) \right)$ with the decomposition 
\begin{equation}\label{eq:reg_energy_raw}
    \calG\left(\bfsigma, \bfvarphi, \nabla\bfvarphi\right)  = \tilde{\calG}\left(\bfsigma, \bfvarphi \right) - \sum_{\alpha=1}^m \sum_{\beta>\alpha} \gamma_\ab \nabla \varphi_\alpha \cdot \nabla \varphi_\beta,
\end{equation}
where $\gamma_\ab$ is a numerical parameter associated with the interface region of $\varphi_\alpha$ and $\varphi_\beta$, which follows \citet{steinbach1999generalized} and has units of an interface energy. The term $\tilde\calG(\bfsigma, \bfvarphi)$ includes the bulk energies $\calG^{\mathrm{bulk}}$ associated with each phase variant and the nonconvex potential energy $\Phi$. It can be written as 
\begin{equation}
    \tilde\calG(\bfsigma,\bfvarphi) = \calG^{\mathrm{bulk}}(\bfsigma, \bfvarphi) + \eta \,\Phi(\bfvarphi),
\end{equation}
where $\eta>0$ is a numerical parameter. $\tilde\calG$ can be seen as a suitably modified version of the Ginzburg-Landau potential \citep{devonshire1949_Part1, schrade2013physical}.  

The order parameter evolves according to \eqref{eq:evo_phi_pre}, thus transporting the regularized interface forward, which carries a certain amount of energy \citep{alber2016alternative}. The energy flow, transported by the level sets of the regularized interfaces, can be written as
\begin{equation}\label{eq:HeatFlow}
\begin{aligned}
        \boldsymbol{\upsilon}^\alpha &= -\frac{1}{m}\sum_{\beta\neq\alpha}  \bfV_{\beal} \gamma_{\alpha \beta} \lvert  \nabla \Psi_{\alpha \beta} \rvert^2, \\
        &=  -\frac{1}{m}\sum_{\beta\neq\alpha} \gamma_{\alpha \beta} V_{\beal}
        \frac{\nabla \Psi_{\alpha \beta}}{\lvert \nabla \Psi_{\alpha \beta} \rvert}
         \lvert  \nabla \Psi_{\alpha \beta} \rvert^2,
\end{aligned}
\end{equation}
where $\bfV_{\beal} = V_{\beal} \bfn$ is the normal velocity vector in the direction of the normal of the level sets.
Note that this energy flow is only related to the motion of the regularized interface; any other thermodynamic energy flow may exist in addition. 
Applying the divergence theorem and inserting the evolution law \eqref{eq:evo_Psi}, the energy flux carried by the regularized interface through the boundary $\pl\calR$ of any bounded subregion $\calR\subseteq\Omega$ is given by 
\begin{equation}
\begin{aligned}
    \int_{\partial \calR} \boldsymbol{\upsilon}^\alpha \cdot \tilde\bfn \dd A %
    &= -\frac{1}{m}\sum_{\beta\neq\alpha} \int_{\partial \calR}\gamma_{\alpha \beta} V_{\beal}  \lvert \nabla \Psi_{\alpha \beta} \rvert  \left(\nabla \Psi_{\alpha \beta} \cdot \tilde\bfn \right) \dd A \\
    &= \int_{\calR} - \frac{1}{m}\sum_{\beta\neq\alpha}  \gamma_{\alpha \beta} \nabla \cdot \left(\parT\Psi_{\alpha \beta} \nabla \Psi_{\alpha \beta} \right) \dd V,
    \end{aligned}
\end{equation}
where $\tilde\bfn$ is the outer normal vector of $\pl\calR$. The first law of thermodynamics, with the account of linear momentum balance and the assumption of no heat sources in the bulk, yields 
\begin{equation}\label{eq:1stlaw}
      \int_{\calR} \bigg[\dot{\calE}  %
     - \nabla \parT\bfu \cdot \bfsigma %
     - \frac{1}{m}\sum_{\beta\neq\alpha}  \gamma_{\alpha \beta} \nabla \cdot \left(\parT\Psi_{\alpha \beta} \nabla \Psi_{\alpha \beta} \right)  %
      + \nabla \cdot \bfq \bigg]\dd V =0,
\end{equation}
where $\calE$ denotes the internal energy density, $\dot{\calE}$ denotes the material derivative, and $\bfq$ the total heat flux density. Solving \eqref{eq:1stlaw} for $\int_\calR\nabla \cdot \bfq\dd V$ and inserting the result into the Clausius-Duhem inequality under isothermal conditions transforms the latter into
\begin{equation}\label{eq:IntermediateInequality}
    \int_{\calR} \bigg[ \dot{\calS}  + \frac{1}{\theta}\left(\nabla \parT\bfu \cdot \bfsigma + \frac{1}{m}\sum_{\beta\neq\alpha}  \gamma_{\alpha \beta} \nabla \cdot \left(\parT\psi_{\alpha \beta} \nabla \psi_{\alpha \beta} \right) - \dot{\calE} \right) \bigg] \dd V \geq 0,
\end{equation}
where $\calS$ denotes the entropy density and $\theta$ is the constant temperature. Since we use a stress-driven formulation of the free energy density in this study, we apply the Legendre transform, $\check{\calG} = \inf_{\bfepsilon} \inf_{\calS} \{ \calE - \theta \calS - \bfepsilon \cdot \bfsigma\}$, which gives 
\begin{equation}
 \dot{\calS}   = \frac{1}{\theta} \left(\dot{\calE}  - \parT\bfepsilon \cdot \bfsigma - \bfepsilon \cdot \parT\bfsigma  - \dot{\check{\calG}}\right),
\end{equation}
where $\parT \bfepsilon \cdot \bfsigma = \nabla \parT \bfu \cdot \bfsigma$ due to the symmetry of $\bfsigma$. Therefore, inequality~\eqref{eq:IntermediateInequality} at a given temperature becomes
\begin{equation}\label{eq:ClausiusDuhem}
   \int_{\calR} \left(- \bfepsilon \cdot \parT\bfsigma + \frac{1}{m}\sum_{\beta\neq\alpha}  \gamma_{\alpha \beta} \nabla \cdot \left(\parT\psi_{\alpha \beta} \nabla \psi_{\alpha \beta} \right) - \dot{\check{\calG}} \right) \dd V \geq 0,
\end{equation}
where $\check{\calG}\left(\bfsigma, \bfPsi, \nabla \bfPsi\right)$ is the regularized free energy density written in terms of the level set functions. Instead of $\check{\calG}$, we prefer to work with the energy density of the order parameters, as it has a cleaner physical meaning. Using the inverse transformation \eqref{eq:inverseTrans_phi} and the relations 
\begin{equation}
     \frac{\partial \check{\calG}}{\partial \Psi_{\alpha \beta}} =\frac{1}{m} \left(\frac{\partial \calG}{\partial \varphi_\alpha} - \frac{\partial \calG}{\partial \varphi_\beta}\right), \quad \frac{\partial \check{\calG}}{\partial \nabla \Psi_{\alpha \beta}} =\frac{1}{m} \left(\frac{\partial \calG}{\partial \nabla\varphi_\alpha} - \frac{\partial \calG}{\partial \nabla\varphi_\beta}\right), 
\end{equation}
we rewrite the Clausius-Duhem inequality \eqref{eq:ClausiusDuhem} as 
\begin{equation}
    \begin{aligned}
         \int_{\calR} \Bigg(- \bfepsilon \cdot \parT\bfsigma + \frac{1}{m} \sum_{\beta\neq\alpha}\bigg[ \gamma_{\alpha \beta}\parT\Psi_{\alpha \beta} \left(\nabla^2 \varphi_\alpha - \nabla^2 \varphi_\beta\right) %
         +  \gamma_{\alpha \beta}\left(\nabla\parT\Psi_{\alpha \beta}\right) \left(\nabla \varphi_\alpha - \nabla \varphi_\beta\right) \\
          - \parT\Psi_{\alpha \beta} \left(\frac{\partial \calG}{\partial \varphi_\alpha} - \frac{\partial \calG}{\partial \varphi_\beta}\right) %
          - \left(\nabla\parT\Psi_{\alpha \beta}\right)\left(\frac{\partial \calG}{\partial \nabla\varphi_\alpha} - \frac{\partial \calG}{\partial \nabla\varphi_\beta}\right) \bigg] - \fp{\calG}{\bfsigma} \cdot \parT\bfsigma \Bigg)\dd V \geq 0.
    \end{aligned}
\end{equation}
When applying the constitutive relation $\partial \calG/\partial \bfsigma = -\bfepsilon$, as in \eqref{eq:Constitutive_StressDriv}$_1$, the inequality becomes 
\begin{equation}
    \begin{aligned}
         \int_{\calR} \frac{1}{m} \sum_{\beta\neq\alpha}\Bigg[ \gamma_{\alpha \beta}\parT\Psi_{\alpha \beta} \left(\nabla^2 \varphi_\alpha - \nabla^2 \varphi_\beta\right) %
         +  \gamma_{\alpha \beta}\left(\nabla\parT\Psi_{\alpha \beta}\right) \left(\nabla \varphi_\alpha - \nabla \varphi_\beta\right) \\
          - \parT\Psi_{\alpha \beta} \left(\frac{\partial \calG}{\partial \varphi_\alpha} - \frac{\partial \calG}{\partial \varphi_\beta}\right) %
          - \left(\nabla\parT\Psi_{\alpha \beta}\right)\left(\frac{\partial \calG}{\partial \nabla\varphi_\alpha} - \frac{\partial \calG}{\partial \nabla\varphi_\beta}\right) \Bigg] \dd V\geq 0.
    \end{aligned}
\end{equation}
Substituting the regularized free energy density~\eqref{eq:reg_energy_raw}, we further arrive at 
\begin{equation}
    \int_{\calR} \frac{1}{m} \sum_{\beta\neq\alpha} \parT\Psi_{\alpha \beta}\left[%
    \gamma_{\alpha \beta}\left(\nabla^2 \varphi_\alpha - \nabla^2\varphi_\beta\right) %
    - \left(\frac{\partial \tilde\calG}{\partial \varphi_\alpha} - \frac{\partial \tilde\calG}{\partial \varphi_\beta}\right) \right] \dd V \geq 0.
\end{equation}
Finally, using the evolution law \eqref{eq:evo_Psi} and the definition of the variational derivative, the Clausius-Duhem inequality reduces to
\begin{equation}
 \int_{\calR}\sum_{\beta\neq\alpha} V_{\beal} \lvert \nabla \Psi_{\alpha \beta} \rvert\left[ %
\frac{\delta \calG}{\delta \varphi_\beta} - \frac{\delta \calG}{\delta \varphi_\alpha}%
    \right] \dd V  \geq 0. 
\end{equation}
This inequality entails a functional relation between $V_\beal$ and the term in the square brackets. In analogy to the sharp-interface model \citep{abeyaratne2006evolution} and the level-set method \citep{hou1999level_Rosakis}, we treat this functional relation as the regularized version of \eqref{eq:CD_sharp} with a driving force that physically drives the level sets of the regularized region with a propagation velocity $V_\beal$. Therefore, we define the regularized driving force as
\begin{equation}\label{eq:reg_drivingforce}
    f_\beal = \frac{\delta \calG}{\delta \varphi_\beta} - \frac{\delta \calG}{\delta \varphi_\alpha},
\end{equation}
and the kinetic relation $\check{V}_\beal(f_\beal)$ has a direct connection to the kinetic relation in the sharp-interface model, as will be shown in the numerical examples in \autoref{sec:results}. The electro-mechanical coupling in the bulk can be incorporated into the model by substituting the free energy density~$\calG$ in \eqref{eq:reg_drivingforce} by its electro-mechanically coupled version, as shown in \citet{cheng2025fft}.



\subsection{The construction of phase fields and its specialization to ferroelectrics}

Ferroelectric materials are characterized by anisotropic electromechanical responses. For the tetragonal crystal system with space group $P4mm$, the material tensors are transversely isotropic with their principal axes aligning with the $c$-axis of the atomic unit cell. The regularized energy density $\calG$ is decomposed into three components: the bulk energy $\calG^\mathrm{bulk}$, the nonconvex potential energy $\Phi$, and the regularizing gradient energy $\calG^\mathrm{grad}$. Based on our stress-driven formulation within each domain with the free energy in~\eqref{eq:gibbs_domain}, the regularized energy density is given by 
\begin{equation}\label{eq:phasefield_general}
\begin{aligned}
    \calG(\bfsigma, \bfe, \varphi, \nabla\varphi ) = %
    \sum_{\alpha=1}^M  \varphi_\alpha \calG_\alpha  + \eta\, \Phi(\bfvarphi) - \sum_{\alpha=1}^M \sum_{\beta>\alpha}\gamma_\ab \nabla\varphi_\alpha \cdot \nabla\varphi_\beta,
\end{aligned}
\end{equation}
where $\eta >0$ is a coefficient of the potential with units of the energy density. In this study, we use a higher-variant version of the multi-obstacle potential \citep{blowey1993curvature_ob_pot,nestler2005multicomponent, Guin2023phase} for the potential energy with the specific form
\begin{equation} \label{eq:ob_pot}
    \Phi(\bfvarphi) =  \sum_{\alpha=1}^M\sum_{\beta >\alpha} 4 |\varphi_\alpha||\varphi_\beta| + \tau \sum_{\alpha=1}^M \sum_{\gamma >\beta > \alpha} \lvert\varphi_\alpha\rvert \lvert\varphi_\beta\rvert \lvert\varphi_\gamma\rvert,
\end{equation}
where $\tau >0$ is the coefficient of the triple-phase term.

For convenience, the double-phase term is normalized such that the 2D slice of the potential, e.g., $\Phi(\varphi, 1-\varphi, 0, 0, 0, 0)$ has a maximum of $1$ at $\varphi = 1/2$. The addition of the triple-variant term avoids the emergence of a spurious third phase in any two-phase interface. The multi-obstacle potential shows certain advantages over the basic second-order multi-well potential, including a finite width of the interface transition region and a more robust enforcement of minima at zero and one compared to the classical double-well potential \citep{blowey1993curvature_ob_pot, blowey1994phase}. (This is in fact visible from their equilibrium profiles: the sinusoidal profile resulting from the obstacle-type potential spreads the transition to the bulk value over a finite region, while the hyperbolic tangent profile resulting from the double-well potential leads to only asymptotic convergence.) These advantages lead to a more accurate evolution of the phase-field profile \citep{garcke1999multiphaseConcept}. Of course, this choice of the potential is not unique and other formulations may be used as well \citep{sakane2026hybrid}. However, it usually requires careful observation to demonstrate all effects of a given potential type.

We interpolate the free energy density of each phase variant, given by \eqref{eq:gibbs_domain}, through the order parameters $\bfvarphi$ to construct the regularized free energy of the bulk in \eqref{eq:phasefield_general}. Alternatively, the regularized energy has also been formulated by interpolating the material tensors of the individual variants%
\footnote{An alternative approach is to construct the regularized material tensors by using the spontaneous polarization of each phase variant \citep{Guin2023phase}. This approach is based on aligning the principal axes of the tensors with the unit cell's crystallographic axes. The rotation of the tensors then follows the rotation of the spontaneous polarization vector. 
However, using this approach, we observe undesirable material constants inside the interpolated regions \citep{schrade2014}. This issue is avoided by constructing the regularization directly with the order parameters $\bfvarphi$. Further details are discussed in \ref{apx:reg_tensor}.} 
\citep{fried1994dynamic_gurtin}, e.g. $\bfEpsS(\bfx, t) = \sum_\alpha^M \varphi_\alpha(\bfx, t) \bfEpsS_\alpha$.
The choice between interpolating the free energy density and interpolating the material tensors can be ambiguous \citep{hou1999level_Rosakis} and in strain-driven formulations may lead to distinct models. This ambiguity originates from the quadratic contributions of the spontaneous strains to the elastic energy. Consequently, when the material tensors are interpolated, the regularized driving force exhibits an explicit dependence on the regularization function; by contrast, this dependence vanishes when the free energy density is interpolated \citep{cheng2025fft}.
Here, however, this ambiguity is avoided by using the stress-based free energy density in \eqref{eq:gibbs_domain}, in which no quadratic terms of a material tensor are involved. This allows us to arrive at the same unique regularized driving force \eqref{eq:reg_drivingforce}, consistent with the discussion in the level-set context \citep{cheng2025fft}. 

\paragraph{The evolution law in ferroelectrics} With the phase-field model discussed above, we arrive at the evolution law specialized to ferroelectrics as 
\begin{equation}\label{eq:evo_phi}
    \parT\varphi_\alpha = \frac{1}{m} 
    \sum_{\beta = 1}^{m} \lvert \nabla\!\left(\varphi_\alpha - \varphi_\beta\right)\rvert \,\check{V}_{\beal}(f_\beal)  
    ,
\end{equation}
where the desired kinetic relation is prescribed directly through the relation $\check{V}_\beal(f_\beal)$%
. 
The regularized driving force is 
\begin{equation} \label{eq:drivingforce_ferro}
    \begin{aligned}
        f_\beal = \left(\calG_\beta - \calG_\alpha \right)  - \eta \left(\frac{\partial\Phi(\bfvarphi)}{\partial \varphi_\alpha} - \frac{\partial \Phi(\bfvarphi)}{\partial \varphi_\beta} \right)  + \gamma_\ab (\nabla^2 \varphi_\alpha - \nabla^2 \varphi_\beta),
    \end{aligned}
\end{equation}
where the difference in the bulk energy of each phase variant is given by 
\begin{equation}
    \begin{aligned}
       \calG_\beta - \calG_\alpha =  %
       &\half \bfe \cdot (\bfEpsS_\alpha - \bfEpsS_\beta) \bfe + (\bfp^s_\alpha - \bfp^s_\beta) \cdot \bfe \\
    &+ \half\bfsigma\cdot(\sfbS_\alpha - \sfbS_\beta) \bfsigma + \bfsigma\cdot(\bfepsilon^s_\alpha - \bfepsilon^s_\beta) +\bfsigma\cdot(\sfbD_\alpha\T - \sfbD_\beta\T )\bfe.
    \end{aligned}
\end{equation}
We note that the derived evolution law \eqref{eq:evo_phi} differs from that in \citet{Guin2023phase} in the prefactor $1/m~\lvert \nabla\!\left(\varphi_\alpha - \varphi_\beta\right)\rvert$. The prefactor $\sqrt{\lvert\nabla \varphi_\alpha \cdot \nabla\varphi_\beta\rvert}$ used by \citep{Guin2023phase} can be viewed as a simplification that is mainly applicable when only a small number of order parameters are present \citep{grose2022multi}. In fact, for a two-phase interface, the proposed model is consistent with the phase-field models in \citet{alber2016alternative}, \citet{agrawal2015dynamic}, and \citet{Guin2023phase}. To demonstrate this, we write $\varphi_\alpha = \varphi$ and $\varphi_\beta = 1 - \varphi$ using the constraint \eqref{eq:constraint_phi}. For this two-phase scenario, \eqref{eq:evo_phi} reduces to $\parT\varphi = \check{V}_\beal(f_\beal)\lvert \nabla\varphi \rvert$, which has mixed hyperbolic--parabolic properties. This two-phase evolution law also shares similar properties with the level-set method \citep{osher1988} under the assumptions discussed in \ref{apx:lsm_gkm}.
 
The coefficients in the phase-field free energy density \eqref{eq:phasefield_general} are connected to physical quantities, including the interface (domain wall) thickness $l$ and the interface energy $\Gamma$. This connection is obtained through the analysis of two-phase interface evolution in 1D, written as the evolution of the order parameter $\varphi$ and solved using the ansatz of a traveling-wave solution \citep{fried1994dynamic_gurtin, Guin2023phase}. Performing the analogous derivation with the multi-obstacle potential \eqref{eq:ob_pot}, we obtain the equilibrium phase profile 
\begin{equation} \label{eq:equiProfile}
    \check{\varphi} =\ds \begin{cases}
        1 \qquad &\text{for} ~(x-Vt) \geq \frac{\pi l}{4} \\
        \frac{1}{2} \left[ 1 + \sin\left({\frac{2}{l} \left(x - Vt\right)}\right)\right]&\text{for} ~ -\frac{\pi l}{4} \leq (x-Vt) \leq \frac{\pi l}{4} \\
        0 &\text{for} ~(x-Vt) \leq -\frac{\pi l}{4} 
    \end{cases},
\end{equation}
where $V$ is the propagation velocity. This profile agrees with the ones reported in \citep{steinbach2023lectures, Guin2023phase}. The interface thickness and energy are obtained from the above as 
\begin{equation} \label{eq:phasefield_relation}
    l =  \sqrt{\frac{\gamma_\ab}{\eta}}, \qquad \Gamma = I \sqrt{\eta \gamma_\ab}
\end{equation}
where $I = 2\int_0^1 \sqrt{\Phi(\check{\varphi})} \dd \check{\varphi}$ is a constant that is specific to a given potential \citep{Plapp2012} and in our case evaluates to $I = \pi /2$. 

In the following, we focus on a single-crystalline ferroelectric whose crystallographic axes are aligned with the global axes $\{\hat\bfe_1, \hat\bfe_2, \hat\bfe_3\}$ in Cartesian coordinates. Under this assumption, the spontaneous polarizations of the material are given as $\bfp^s_1 = p_0 \hat{\bfe}_1$, $\bfp^s_2 = - p_0 \hat{\bfe}_1$, $\bfp^s_3 = p_0 \hat{\bfe}_2$, $\bfp^s_4 = - p_0 \hat{\bfe}_2$, $\bfp^s_5 = p_0 \hat{\bfe}_3$, and $\bfp^s_6 = - p_0 \hat{\bfe}_3$, where $p_0$ is the spontaneous polarization magnitude of the material at the prescribed temperature. Examples of spontaneous polarizations are illustrated in \autoref{fig:dw_180_90}a. The ferroelectric material of $P4mm$ symmetry exhibits two classes of domain walls: 90$\degree$ and 180$\degree$. Introducing the set of all order parameter pairs as $\calP = \{\{\varphi_\alpha, \varphi_\beta\}, | \alpha \neq \beta\}$, the two domain wall types motivate the definition of two sets of order parameter pairs: $\calP_{180} = \{\{\varphi_1, \varphi_2\}, \{\varphi_3, \varphi_4\}, \{\varphi_5, \varphi_6\}\}$, and $\calP_{90} = \calP\setminus\calP_{180}$. This allows us to define the phase-field model parameters independently for the two sets relating to the domain wall type. For example, we can simplify the gradient coefficients $\gamma_\ab$ to two numbers by writing $\gamma_{180} = \gamma_{0}$ and $\gamma_{90} =\gamma_{0} g$, where $g = \gamma_{90}/\gamma_{180}$. This will also simplifies the nondimensionalization process below. 

\subsection{Nondimensionalization}\label{sec:nondim}

We follow \citet{cheng2025fft} to perform nondimensionalization of the governing equations in this study. This not only helps the convergence of the spectral solver in our simulations, but it also helps to identify the physical implications of the parameters used in the phase-field model. The material properties and the field quantities are nondimensionalized as 
\begin{equation} \label{eq:nonDimension_FFT}
\begin{aligned}
        &\ol \bfp^s = \frac{\bfp^s} {p_0}, \quad \ol \bfe = \frac{\bfe}{p_0/\EpsS_{c}}, \quad \overline{\bfEpsS} = \frac{\bfEpsS}{\EpsS_c} , \quad \ol \bfd = \frac{\bfd}{p_0}, \quad \ol \bfepsilon^s = \frac{\bfepsilon^s}{\varepsilon_c} , \quad \ol \sfbS = \frac{\sfbS}{\EpsS_c\varepsilon_c^2/p_0^2}, \quad \ol \bfsigma = \frac{\bfsigma}{p_0^2/\EpsS_c \varepsilon_c},\\ 
        & \ol\bfepsilon = \frac{\bfepsilon}{\varepsilon_c} ,\quad \ol \sfbD = \frac{\sfbD}{\EpsS_c\varepsilon_c/p_0}, 
\end{aligned}
\end{equation}
where $\varepsilon_c$ is the spontaneous strain along the $c$-axis of the atomic crystal, and $\EpsS_c$ the dielectric permittivity at constant stress also along the $c$-axis. The quantities related to the evolution law are nondimensionalized as
\begin{equation} \label{eq:nonDimension_evoLaw}
\begin{aligned}
        &\ol{f_\beal} = \frac{f_\beal}{p_0^2/\EpsS_{c}}, \quad \ol{\bfx} = \frac{\bfx}{l_0}, \quad \ol{\check{V}_\beal}(\ol{f_\beal}) = \frac{\check{V}_\beal(f_\beal)}{V_0}, \quad \ol{t} = \frac{V_0}{l_0} t,
\end{aligned}
\end{equation}
where $V_0= \check{V}(p_0^2/\EpsS_{c})$ is the characteristic velocity and $l_0 = l/h$ is the characteristic length. Applying these dimensionless quantities, the nondimensionalized driving force is given by 
\begin{equation} \label{eq:drivingforce_dimless}
    \ol{f_\beal} =  \left(\ol{\calG_\beta} - \ol{\calG_\alpha} \right)  + z \left( - \frac{1}{h^2} \left(\frac{\partial\Phi(\bfvarphi)}{\partial \varphi_\alpha} - \frac{\partial \Phi(\bfvarphi)}{\partial \varphi_\beta} \right)  + g (\ol{\nabla}^2 \varphi_\alpha - \ol{\nabla}^2 \varphi_\beta)\right),
\end{equation}
where the relations \eqref{eq:phasefield_relation} between the parameters in phase-field model are used, The dimensionless parameter $z$ in \eqref{eq:drivingforce_dimless} is obtained as 
\begin{equation} \label{eq:reglarizedPar}
    z = \frac{\EpsS_{c} \Gamma h^2}{p_0^2 l I} = \frac{1}{f_0} \frac{\Gamma}{l} \frac{h^2}{I},
\end{equation}
where $f_0 = p_0^2/\EpsS_{c}$ is the characteristic driving force in \eqref{eq:nonDimension_evoLaw}, and $z$ can be considered as the regularization parameter that encodes the length scale of the problem. In the limiting case, the regularized driving force becomes $\lim_{z \rightarrow 0} \ol{f_\beal} = \ol{\calG_\beta} - \ol{\calG_\alpha} $, which resembles the one from the level-set method without any regularization \citep{cheng2025fft}. The evolution law retains its form after nondimensionalization and is given by 
\begin{equation}\label{eq:evo_phi_dimless}
    \ol{\parT}\varphi_\alpha = \frac{1}{m} \left(\sum_{\beta = 1}^{m} \ol{\check{V}_{\beal}}(\ol{f_\beal})  
    ~\lvert \ol{\nabla}\!\left(\varphi_\alpha - \varphi_\beta\right)\rvert
    \right).
\end{equation}

The form of the remaining governing equations, such as the field equations \eqref{eq:fieldBalance} and the free energy density \eqref{eq:gibbs_domain}, remains unchanged after nondimensionalization. In the following, we focus on the model behavior; consequently, the dimensionless equations are applied in the numerical examples of this study, and the overhead bar is dropped for the sake of simplicity in the remainder of the article.  

\subsection{Numerical implementation using an FFT spectral solver}

The response time of the electric and mechanical fields is generally assumed significantly shorter than that of domain evolution in all applications considered here. This motivates a staggered solution scheme \citep{choudhury2005phase,Vidyasagar2017}: at every time step, we first solve the homogenization problem for the current domain configuration to obtain the full-field stress and electric fields, and subsequently evolve the domain walls using the evolution law with the driving force computed from those fields.

The homogenization problem with the balance laws \eqref{eq:fieldBalance} is numerically solved using the FFT-based Galerkin method, as discussed in \citet{cheng2025fft}, to obtain the full-field stress and electric field data for applied average ones. The Fast-Fourier Transform (FFT) enables efficient constructions of finite-dimensional function spaces with trigonometric polynomials for the Galerkin discretization. We define the scaled wave vector $\xi_i = 2 \pi k_i/L$ with  $k_i \in \dsZ$ being the wave vector and $L$ the length of a cuboid-shaped representative volume element (RVE). The variational form of the homogenization problem leads to the equilibirum conditions 
\begin{equation} \label{eq:fftEquilibrium}
     \sfbR \ast \left(\sfbS_\alpha \bfsigma +\bfepsilon^s_\alpha +\sfbD_\alpha\T \bfe\right)=0, \quad \bfG \ast \left( \bfEpsS_\alpha \bfe + \bfp^s_\alpha + \sfbD_\alpha \bfsigma \right)=0,
\end{equation}
where $\ast$ denotes the convolution operation. The equilibrium conditions \eqref{eq:fftEquilibrium} are more easily solved in the Fourier space. The first Green's operator $\sfbR$ in Fourier space has components
\begin{equation} \label{eq:R_operator}
    \hat \sfR_{ijkl} =  
    \begin{cases}
       \ds \hat{\sfI}_{ijkl} - %
       \frac{1}{2\lvert\bfxi\rvert^2} \left( \delta_{ik} \xi_l \xi_j + \delta_{il} \xi_k \xi_j + \delta_{kj} \xi_l \xi_i + \delta_{lj} \xi_i \xi_k - 2 \frac{\xi_i \xi_j \xi_k \xi_l}{\lvert\bfxi\rvert^2} \right) \quad &\bfxi \neq  \bfnull, \\
        0 \quad &\bfxi = \bfnull,
    \end{cases}
\end{equation}
where $\hat{\sfI}_{ijkl}=1/2(\delta_{ik} \delta_{jl} + \delta_{il} \delta_{jk})$ are the components of the fourth-order identity tensor. The second Green's operator $\bfG$ in Fourier space has components
\begin{equation}\label{eq:GreenOperator_G}
       \hat G_{ij}(\bfxi) = 
    \begin{cases}
   \ds \frac{\xi_i \xi_j}{\lvert\bfxi\rvert^2} \quad &\bfxi \neq  \bfnull,\\[10pt]
    0 \quad &\bfxi = \bfnull.
\end{cases}
\end{equation}
We apply Newton-Raphson iteration to the equilibrium conditions \eqref{eq:fftEquilibrium} with the assumption of small increments between iterations:
\begin{equation}
    \bfsigma^{(i+1)} = \bfsigma^{(i)} + \Delta \bfsigma^{(i)}, \quad \bfe^{(i+1)} = \bfe^{(i)} + \Delta \bfe^{(i)},
\end{equation}
where $i$ denotes the $i$-th iteration. Since the constitutive response is linear, the Newton-Raphson scheme reduces to the system of equations 
\begin{equation}
    \begin{aligned} \label{eq:lysNR_stressDriven}
        \invFT{\hat\sfbR \fwdFT{\sfbS\Delta\bfsigma\nstp + \sfbD\T \Delta \bfe \nstp}}&= -\invFT{\hat\sfbR\fwdFT{\sfbS\bfsigma\nstp + \sfbD\T\bfe\nstp + \bfepsilon^s}}\\
        \invFT{\hat\bfG\fwdFT{\bfEpsS\Delta\bfe\nstp+\sfbD\Delta\bfsigma\nstp}} &= -\invFT{\hat\bfG \fwdFT{\bfEpsS\bfe\nstp+\sfbD\bfsigma\nstp+\bfp^s}}, 
    \end{aligned}
\end{equation}
where we use $\invFT{\cdot}$ and $\fwdFT{\cdot}$ to denote the inverse and forward Fourier transform, respectively. In each Newton-Raphson iteration, the preconditioned conjugate gradient (PCG) method is used to solve this system of equations. The averages are imposed through the initial guesses
\begin{equation}
    \bfsigma^{(0)} = \ol \bfsigma, \quad \bfe^{(0)} = \ol \bfe.
\end{equation}
The Newton-Raphson scheme converges when the right-hand side of \eqref{eq:lysNR_stressDriven} approaches zero, indicating that the equilibrium conditions \eqref{eq:fftEquilibrium} have been satisfied. We note that the superscript $\bfsigma^{(i)}$ denotes the $i$-th Newton-Rapshon iteration step (and must be distinguished from the time step notation described in the following). 

After obtaining the full-field quantities, the evolution law \eqref{eq:evo_phi} is solved using an explicit scheme, following \citet{Vidyasagar2017}, so that the value of $\varphi_\alpha$ at time $t^{(n+1)} = (n+1) \Delta t$ is given as
\begin{equation}
    \varphi_\alpha^{(n+1)} = \varphi_\alpha^{(n)} + \frac{\Delta t}{m} \left(\sum_{\beta = 1}^{m} \check
    {V}_{\beal}^{(n)}(f_\beal^{(n)})  ~\lvert \nabla\!\left(\varphi_\alpha^{(n)} - \varphi_\beta^{(n)}\right)\rvert \right),
\end{equation}
in which the Laplacian term in the driving force \eqref{eq:drivingforce_dimless} and the norm \eqref{eq:evo_phi} are calculated in Fourier space and projected back to real space. GPU-accelerated computation is used in this work and becomes effective due to the adoption of the FFT spectral method in both the full-field calculations and evolution laws.

We implement the general kinetic model on a two-dimensional (2D) RVE with a regular grid of $N_1\times N_3$ grid points, which consists of $N_{l}$ points in the regularized domain wall region ($N_{l}=4$ to $6$ grid points is recommended). This range of points in the regularized region is typical for phase-field models and represents a balance between numerical cost and accuracy. Periodicity conditions are applied at the boundaries. The dimensionless governing equations are used in the following results with the nondimensionalization provided in \autoref{sec:nondim}.





\section{Results} \label{sec:results}

We focus on single-crystalline ferroelectric materials that exhibit a tetragonal perovskite structure at room temperature. We select barium titanate (BaTiO$_3$) as the representative material with properties listed in \ref{apx:Mat_BTO}. BaTiO$_3$ is one of the classic tetragonal ferroelectrics \citep{Devonshire1954,lines2001principles, meier2020domain} and shows a highly anisotropic electromechanical behavior \citep{cheng2025fft}. 


\subsection{Nonlinear kinetic relations}

We begin by demonstrating the capability of the general kinetic model introduced above to assign arbitrary admissible kinetic relations $\check{V}_\beal(f_\beal)$ to simulated interfaces through the evolution laws \eqref{eq:evo_phi}. As an example scenario, we choose a straight 180$\degree$ charge-neutral domain wall. This case is especially relevant in the evolution of ferroelectric domains, since nonlinear relations have indeed been reported for the kinetic relation of this type of domain wall in a purely electrostatic approximation \citep{miller1958velocity, liu2016intrinsic}. To further reduce complexity and to cleanly demonstrate the recovery of the exact kinetic relation, we specialize the problem to anisotropic electrostatics, so that the bulk (sharp-interface) driving force remains constant during domain evolution when subjected to an electric field along the $\hat{\bfe}_3$-direction, i.e., $\bfe = (0, \ol e)\T$. In this case, the dimensionless sharp-interface driving force \eqref{eq:SI_DF_stressDriven} simplifies to $\check{f}= \langle \bfe \rangle \cdot \llbracket \bfp^s \rrbracket =2\ol e$. In particular, this allows us to apply the Merz--Stadler law 
\citep{Merz1954, miller1958velocity, miller1959, stadler1966forward} (summarized and reformulated with the driving force in \citet{Guin2023phase} for pure electrostatics) as one of the two representative nonlinear kinetic relations in our examples.

The first kinetic relation tested (Merz--Stadler law) exhibits an inverse exponential law below a threshold driving force $\tilde f_t$ and a power law above the threshold value, as introduced in \autoref{sec:intro}. These two regimes can be interpreted, respectively, as thermally assisted pinning/depinning domain wall motion and a strong-driving-force regime in which the wall moves immediately in response to an applied field. This kinetic relation takes the form 
\begin{equation}\label{eq:KR_expPow}
    \check{V}^{(1)} = \begin{cases}
        \ds \sign{(f)}V_l \exp\!\left(\frac{-f_a}{\lvert f \rvert}\right) &\quad \lvert f \rvert \leq \tilde{f}_t,\\[15pt]
        \ds \sign{(f)}V_h \left(\frac{-f_a}{\lvert f \rvert}\right)^{r}&\quad \lvert f \rvert > \tilde{f}_t,
    \end{cases}
\end{equation}
where $f_a>0$ is the activation driving force, and $V_l,V_h>0$ are the characteristic velocities of the low- and high-driving-force regions, respectively. In our simulations, these parameters are chosen as $r = 1.4$, $\tilde{f}_t = 1.2$,  $f_a = 3.6$, $V_l = 25$, and $V_h = V_0 \exp\!\left(-f_a / \tilde{f}_t\right)$ for continuity. 

The second kinetic relation tested is of threshold kinetic type, with a square root relation emerging once the driving force exceeds a critical threshold magnitude. This relation was suggested to resemble the phenomenon of phase boundary motion, which is characterized by stick-slip-like behavior of overcoming periodic local energy minima \citep{abeyaratne1996kinetics, abeyaratne1997kinetics, lecomte2009depinning, jo2009nonlinear, tian2025intrinsic}. This kinetic relation is given by 
\begin{equation}\label{eq:KR_sqrt}
    \check{V}^{(2)} = \begin{cases}
        \sign{(f)}\sqrt{f^2 - f_t^2} &\quad \lvert f \rvert > f_t, \\
        0 &\quad \text{else~},
    \end{cases}
\end{equation}
where $f_t = 0.8$ is the chosen threshold of the driving force in simulations. 

In simulations, we use a grid size of $600\times12$ with 6 points in the regularized domain wall. The time step $\Delta t$ is $5\times10^{-4}$. The regularized driving force in this specialized electrostatic case is given by 
\begin{equation}\label{eq:drivingforce_ferro_elec}
    \begin{aligned}
        f_\beal = & \half \bfe \cdot (\bfEpsS_\alpha - \bfEpsS_\beta) \bfe + (\bfp^s_\alpha - \bfp^s_\beta) \cdot \bfe \\
        &+ z \left[ - \frac{1}{h^2} \left(\frac{\partial\Phi(\bfvarphi)}{\partial \varphi_\alpha} - \frac{\partial \Phi(\bfvarphi)}{\partial \varphi_\beta} \right)  + g \left(\nabla^2 \varphi_\alpha - \nabla^2 \varphi_\beta\right)\right],
    \end{aligned}
\end{equation}
and the dimensionless material parameters are chosen as $z = 0.7$, $g=h=1$. 

\begin{figure}
    \centering
    \includegraphics[width=1\linewidth]{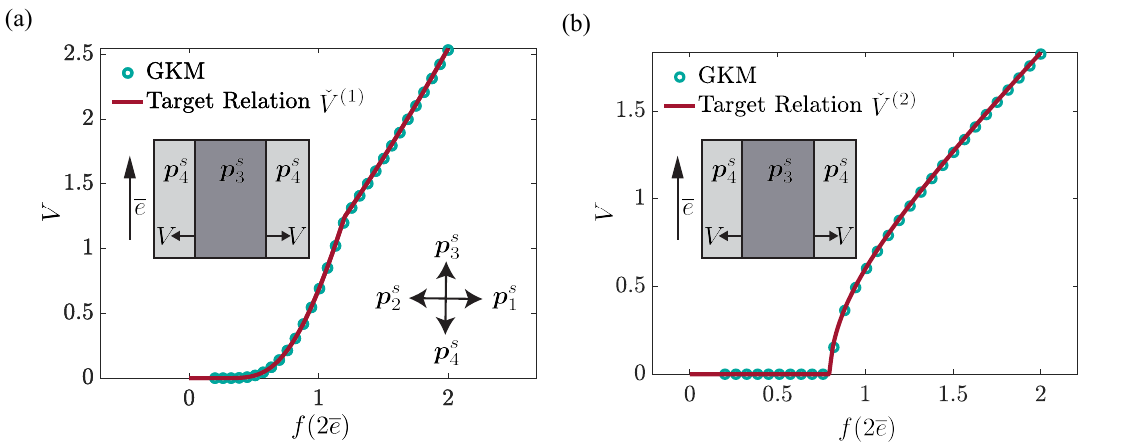}
    \caption{Verification of nonlinear domain wall evolution kinetics in the general kinetic phase-field model (GKM) by simulating $180\degree$ domain walls with prescribed nonlinear kinetic relations: (a)~transition from an exponential law to a power-law according to relation $\check{V}^{(1)}$, (b)~a threshold square-root relation corresponding to $\check{V}^{(2)}$. The target relations are plotted against the sharp-interface driving force $f_s=2\ol{e}$, where $\ol{e}$ is the electric field applied in the simulation. For simplicity, a purely electrostatic model (without piezoelectric coupling) is considered here.}
    \label{fig:kineticRelation}
\end{figure}

The only non-zero order parameters in this case of a straight 180$\degree$ charge-neutral domain wall are $\{\varphi_\alpha, \varphi_\beta\} = \{\varphi_3, \varphi_4\}$. For a constant applied electric field, the driving force remains constant. Hence, the domain wall velocity in the simulation is computed as the displacement of the wall divided by the travel time. We vary the electric field in the range $\ol e\in[0, 1]$ to obtain the domain wall velocity. \autoref{fig:kineticRelation} compares the recorded domain wall velocity from the general kinetic model (GKM) with the target relation evaluated with the driving force of the sharp-interface model: $f_s=2\ol{e}$. \autoref{fig:kineticRelation}a and b show the results obtained for the kinetic relations $\check{V}^{(1)}$ and $\check{V}^{(2)}$, respectively. The close agreement in both cases demonstrates that the model evolves the domain wall accurately according to the prescribed kinetic relation, which is imposed directly through the evolution law \eqref{eq:evo_phi}. These numerical results further confirm the theoretical derivation of the evolution of a straight 180$\degree$ domain wall in \citet{Guin2023phase}.
%

\subsection{Assigning distinct kinetics for different types of domain walls}
In perovskites, the evolution of 180$\degree$ domain walls can be distinct from that of 90$\degree$ ones. This is attributed to the fundamental difference between these two types of domain walls. At a 180$\degree$ domain wall, spontaneous polarization vectors of antiparallel orientation meet, which does not involve a mismatch of the spontaneous strain at the interface. By contrast, a 90$\degree$ domain wall separates two spontaneous polarization vectors that are approximately 90$\degree$ from each other, exhibiting a mismatch in spontaneous strain \citep{damjanovic1998, meyer2002ab}. Therefore, a 90$\degree$ domain wall is thicker and structurally more complex than a 180$\degree$ one \citep{foeth1999comparison, volker2011multiscale}, and is hence more sensitive to mechanical stimuli \citep{tagantsev2010, sumigawa2020situ, ferroelasticMaterial_book_2025}. The GKM introduced here allows us to assign distinct interface properties and kinetic evolution equations to each type of domain wall, unlike in classical Allen-Cahn phase-field models of ferroelectrics \citep{zhang_Bhattacharya_2005,su_landis_2007,indergand2020}.
\begin{figure}[!b]
    \centering
    \includegraphics[width=1\linewidth]{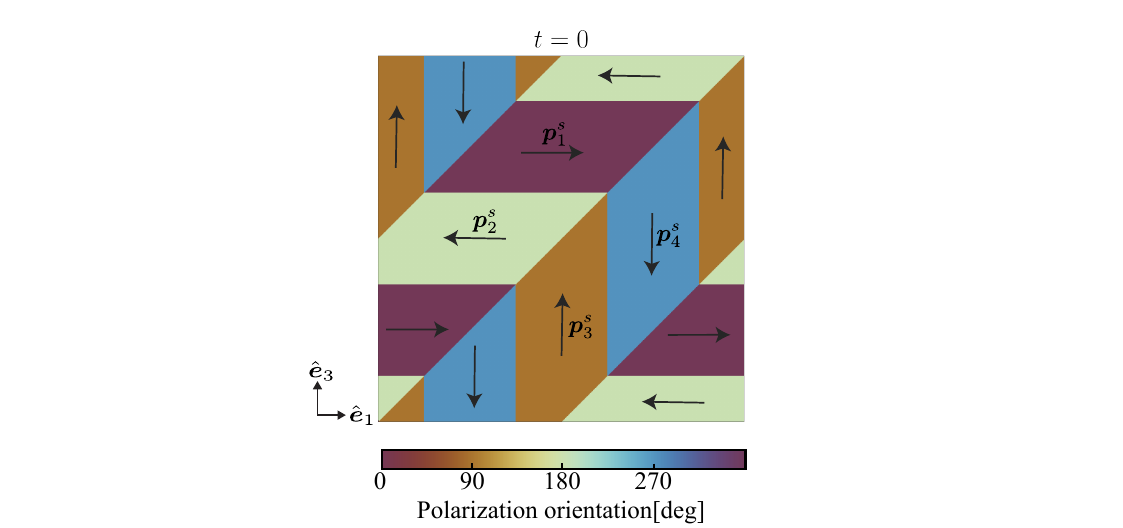}
    \caption{Rank-2 laminate domain structure, which features a combination of $180\degree$ and $90\degree$ charge-neutral domain walls. Spontaneous polarization directions are indicated by black arrows. A similar domain structure was observed by \citet{arlt1990twinning, baither2001ferroelastic}.} 
    \label{fig:initPol}
\end{figure}
As a representative example of a more complex domain configuration, we simulate a rank-2 laminate structure \citep{arlt1990twinning, baither2001ferroelastic}, in which both $180\degree$ and $90\degree$ walls are present, as shown in \autoref{fig:initPol}.%
The square RVE in simulations is constructed with the parameters listed in \autoref{tab:num_par}. 

\begin{table}[!tb]
	\centering
	\caption{Summary of numerical square RVE parameters}
	\label{tab:num_par}
	\begin{tabular}{l l l}
 	\hline
      Parameter& ~ & Value\\
        \hline
     	RVE size ($N_{1} \times N_{3}$)& ~ & $512 \times 512$\\
        No. points in the interface ($N_{l}$) & ~ & 4\\
        $\Delta t$ & ~ & $1\times10^{-3}$\\
        $\Delta x$ & ~ &  1/4
\end{tabular}
\end{table}

\subsubsection{Suppressed kinetics of 90 or 180 degree domain walls}
\label{sec:SuppressedKinetics}

In this example, we demonstrate that prescribing different kinetic relations for $180\degree$ and 90$\degree$ domain walls is straightforward. The parameters of the regularized driving force are set to $z=h=1$ and $g=0.5$. To drive domain-wall motion, we apply an average electric field of magnitude 0.2 in the $\hat{\bfe}_3$-direction, i.e., $\langle \bfe \rangle = (0,0.2)\T$, together with a zero average stress $\langle \bfsigma \rangle = \bfnull$.

\begin{figure}
    \centering
    \includegraphics[width=1\linewidth]{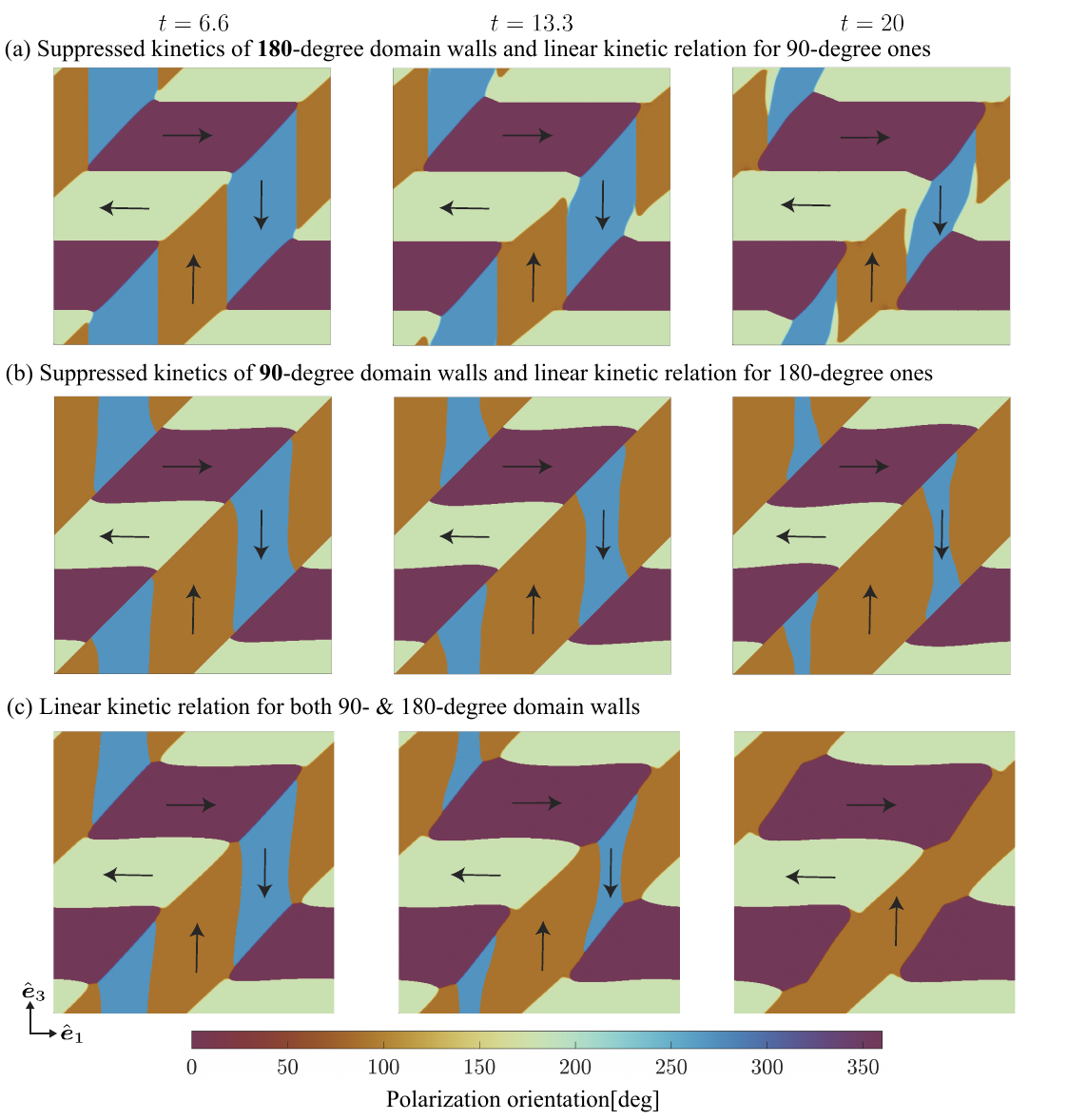}
    \caption{Domain evolution in a rank-2 laminate structure with distinct kinetics prescribed for $180\degree$ and $90\degree$ domain walls. An average electric field $\langle \bfe \rangle = (0,0.2)\T$ and zero average stress are applied to the RVE. (a) A linear kinetic relation with unit mobility is applied to all $90\degree$ domain wall pairs, while a zero velocity is enforced on $180\degree$ domain wall pairs. (b) Conversely, a zero velocity is prescribed on $90\degree$ domain wall pairs, and a linear kinetic relation with unit mobility is applied to $180\degree$ domain wall pairs. (c) A linear kinetic relation with unit mobility is applied to all domain wall pairs.} 
    \label{fig:Pinned_walls}
\end{figure}

To obtain a clear contrast in the resulting microstructural evolution, we selectively suppress either 180$\degree$ or 90$\degree$ domain wall motion by setting the kinetic relation for the corresponding order-parameter pairs to zero. \autoref{fig:Pinned_walls}a shows the domain evolution with $\check{V}_{\beal}(f_\beal)=0$ for $\{\varphi_\alpha,\varphi_\beta\}\in\calP_{180}$ and $\check{V}_{\beal}(f_\beal)=f_\beal$ for $\{\varphi_\alpha,\varphi_\beta\}\in\calP_{90}$ with a linear kinetic relation with unit mobility $\check{V}_\beal(f_\beal) = f_\beal$. In contrast, \autoref{fig:Pinned_walls}b shows the domain evolution with $\check{V}_{\beal}(f_\beal)=0$ for $\{\varphi_\alpha,\varphi_\beta\}\in\calP_{90}$ and $\check{V}_{\beal}(f_\beal)=f_\beal$ for $\{\varphi_\alpha,\varphi_\beta\}\in\calP_{180}$. In both cases, the suppressed kinetics cause the corresponding two-phase domain walls to remain immobile and straight, while inducing distinct motion at domain wall junctions -- which confirms the ability to independently control the kinetics of the two domain wall types.

When the same linear kinetic relation is applied to both types of domain walls, as shown in \autoref{fig:Pinned_walls}c, the junction evolution reflects the combined effect of the two suppressed-kinetics cases. We also note the importance of the parameter $\tau$ in \eqref{eq:ob_pot} for simulations involving more than two phases; throughout this work, we set $\tau=20$. When $\tau=0$, a spurious third phase can emerge within an otherwise two-phase interface \citep{blowey1993curvature_ob_pot, garcke1999multiphaseConcept, nestler2005multicomponent, Guin2023phase}.

More generally, a domain wall junction is a singular point, which is expected to have its own driving force and kinetic relation, distinct from those of a simple two-phase interface \citep{simha1998kinetics}. For ferroelectrics, obtaining both the sharp-interface driving force and the associated kinetic law for such junctions remains an open challenge that we do not address here.

%

\subsubsection{The gradient coefficients for 90 and 180 degree domain walls}
\begin{figure}
    \centering
    \includegraphics[width=1\linewidth]{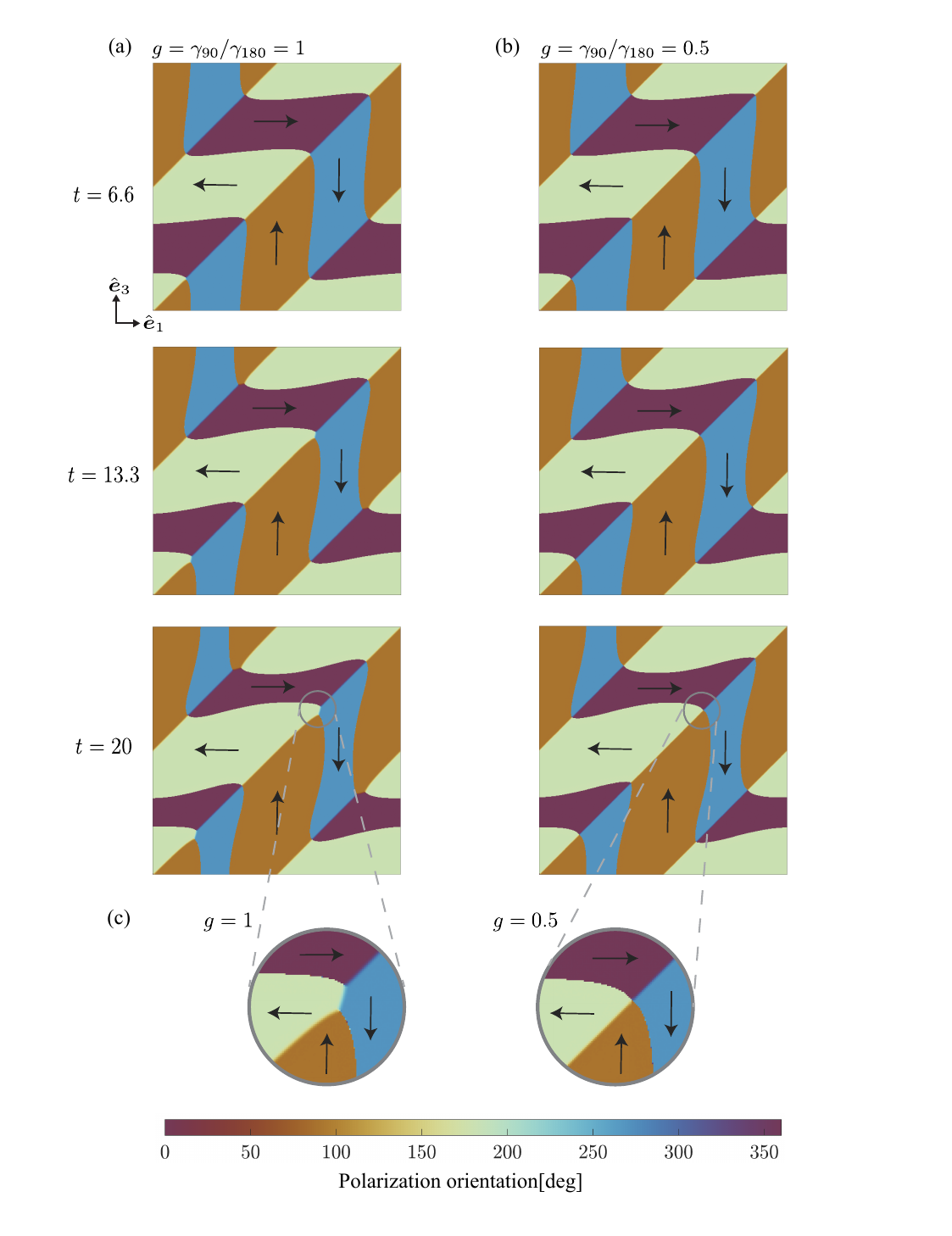}
    \caption{Comparison of domain evolution in a rank-2 laminate structure with uniform versus domain wall type-dependent interface energy. An average electric field $\langle \bfe \rangle =(-0.2/\sqrt{2}, 0.2/\sqrt{2})\T$ and zero average stress are applied in both cases. (a) Domain wall evolution with identical interfacial energies, i.e., $g = \gamma_{90}/ \gamma_{180} =1$. (b) Evolution with distinct interfacial energies with $g=0.5$. Regions highlighted by gray circles in (a,b) are magnified in (c), showing a quadruple junction: with identical interfacial energies (left) we observe junction splitting, whereas distinct interfacial energies (right) retain a stable junction for the prescribed ratio of $g=0.5$.} 
    \label{fig:gCoff}
\end{figure}
Ferroelectrics generally exhibit different interfacial energies of 90$\degree$ and 180$\degree$ domain walls. This motivated introducing the coefficient $\gamma_\ab$ in the regularized energy density \eqref{eq:phasefield_general}, which led to the gradient coefficient $g=\gamma_{90}/\gamma_{180}$ in the regularized driving force \eqref{eq:drivingforce_dimless}. Together with \eqref{eq:phasefield_relation}, this coefficient enables prescribing a distinct physical interface energy for each domain wall type.

Using the same rank-2 laminate structure from above, the effect of introducing the anisotropic gradient coefficient~$g$ is examined in \autoref{fig:gCoff}. An average electric field $\langle \bfe \rangle = (-0.2/\sqrt{2}, 0.2/\sqrt{2})\T$ is applied under a 135$\degree$ angle to the positive $\hat{\bfe}_1$-axis, together with a zero average stress $\langle \bfsigma\rangle = \bfnull$. A linear kinetic relation with unit mobility $\check{V}_\beal(f_\beal) = f_\beal$ is prescribed to all order parameter pairs. As shown in \autoref{fig:gCoff}, this combination of applied fields simultaneously favors the domains with $\bfp^s_2$ and $\bfp^s_3$ and disfavors the domains having $\bfp^s_1$ and $\bfp^s_4$. For $g = 1$ (identical interfacial energies),  \autoref{fig:gCoff}a shows the splitting of quadruple junctions into triple junctions
, as highlighted by a gray circle and the magnification in \autoref{fig:gCoff}c. Similar splitting under an isotropic interface energy condition has been observed in the distinct but related problem of grain growth \citep{fan1997diffusion, cahn1999co,perumal2020quadrijunctions}.

Conversely, for the simulation in \autoref{fig:gCoff}b we set $g=0.5$ -- a typical interface energy ratio for 90$\degree$ to $180\degree$ domain walls for PbTiO$_3$ using the atomistic shell-model potential \citep{volker2011multiscale}. In contrast to the case of identical interfacial energies (splitting at $t=13.3$), \autoref{fig:gCoff}b shows stable quadruple junctions throughout the simulation.

These results demonstrate that introducing domain wall type-dependent interface energy has an essential effect on the evolution of ferroelectric domains when multiphase junctions are involved. Unlike in previous models \citep{Guin2023phase}, the presented general kinetic model (GKM) easily admits assigning distinct values to the different types of interfaces. It should be noted that the presented results are insufficient to draw general conclusions about the stability of singular points such as quadruple junctions. For instance, the quadruple junctions in our simulations do split, if an average electric field is applied under a $90\degree$ angle to the positive $\hat{\bfe}_1$-axis, as displayed in \autoref{fig:Pinned_walls}c. The stability of the junction requires more thorough analysis, which is beyond the scope of this study. However, aside from the ability to assign distinct domain wall kinetics (as discussed in Section~\ref{sec:SuppressedKinetics}), our model admits assigning distinct domain wall energies, which jointly contribute to the domain wall evolution under applied fields.

\subsection{Influence of anisotropy on 90 degree domain evolution in barium titanate}

Having demonstrated how distinct domain wall kinetics can affect the evolution of domain structures, we proceed to highlight a further important ingredient that influences the evolution of 90$^\circ$ domain walls in laminate patterns -- enabled by the new GKM framework. 

Single-crystalline barium titanate exhibits a strongly anisotropic permittivity, quantified by the ratio $\nu = \EpsS_a/\EpsS_c$, where $\EpsS_a$ and $\EpsS_c$ denote the permittivity along the $a$- and $c$-axes of the unit cell, respectively. We use our model to demonstrate that this anisotropy can cause the driving force on a $90\degree$ domain wall to undergo a sign change as the magnitude of the applied electric field increases. To isolate the role of permittivity, we first analyze the purely electrostatic contribution and derive the transition point analytically. The resulting insights can then be carried over to the fully electromechanical setting, as the electrostatic contribution plays a significant role in domain wall evolution \citep{cheng2024growth}.

In this purely electrostatic problem, we consider a straight $90\degree$ neutral domain wall that separates the RVE in half with periodic boundary conditions, as shown in \autoref{fig:dw_180_90}b. We denote the domain with polarization $\bfp^s = (\sqrt{2}, \sqrt{2})\T$ as $\Omega_\alpha$, and the domain with $\bfp^s = (\sqrt{2}, -\sqrt{2})\T$ as $\Omega_\beta$. An electric field of magnitude $\ol e$ is applied under a $45\degree$ angle with the positive $\hat{\bfe}_1$-axis, i.e., $\langle \bfe \rangle = (\ol e/\sqrt{2}, \ol e / \sqrt{2})\T$. Since the electric fields are constant inside each domain, we can solve for these fields from the jump conditions \eqref{eq:jumpCond_elec} and the applied average electric field \citep{cheng2025fft}, which yields 
\begin{equation}
    \bfe^\alpha = \begin{pmatrix}
        \ds \frac{\ol{e} \nu \sqrt{2} }{ 1+ \nu}\\[10pt]
       \ds  \frac{\ol{e}}{\sqrt{2}}
    \end{pmatrix}, %
    \quad  %
    \bfe^\beta = \begin{pmatrix} 
        \ds \frac{\ol{e} \sqrt{2} }{ 1+ \nu}\\[10pt]
        \ds \frac{\ol{e}}{\sqrt{2}}
    \end{pmatrix}.
\end{equation}
With these electric fields, the driving force \eqref{eq:SI_DF_stressDriven} of the sharp-interface model is obtained as
\begin{equation} \label{eq:df90_elec}
    \check{f}(\ol{e}, \nu) = - \frac{\ol{e}}{2} \big[ -2 + \ol{e} \left( \nu - 1 \right)\big]. 
\end{equation}
With increasing electric field, this driving force changes sign at a critical value 
\begin{equation}\label{eq:criticalValue}
    \ol{e}_{\mathrm{ct}} = 2/(\nu -1). 
\end{equation}
This change in the direction of the driving force indicates that, contrary to the situation at low electric fields, the domain $\Omega_\beta$ becomes favorable over $\Omega_\alpha$ at higher applied electric fields. For BaTiO$_3$ with $\nu \approx 26$, the critical value is obtained as $\ol{e}_{\mathrm{ct}} = 0.08$, which is equivalent to $15$~MV/m. (For reference, single-crystal BaTiO$_3$ was reported to exhibit a coercive field of $50$~kV/m \citep{burcsu2004large} and a breakdown field of $45$~MV/m \citep{inuishi1958electric,morrison2005high}.)
\begin{figure}[tb]
    \centering
    \includegraphics[width=1\linewidth]{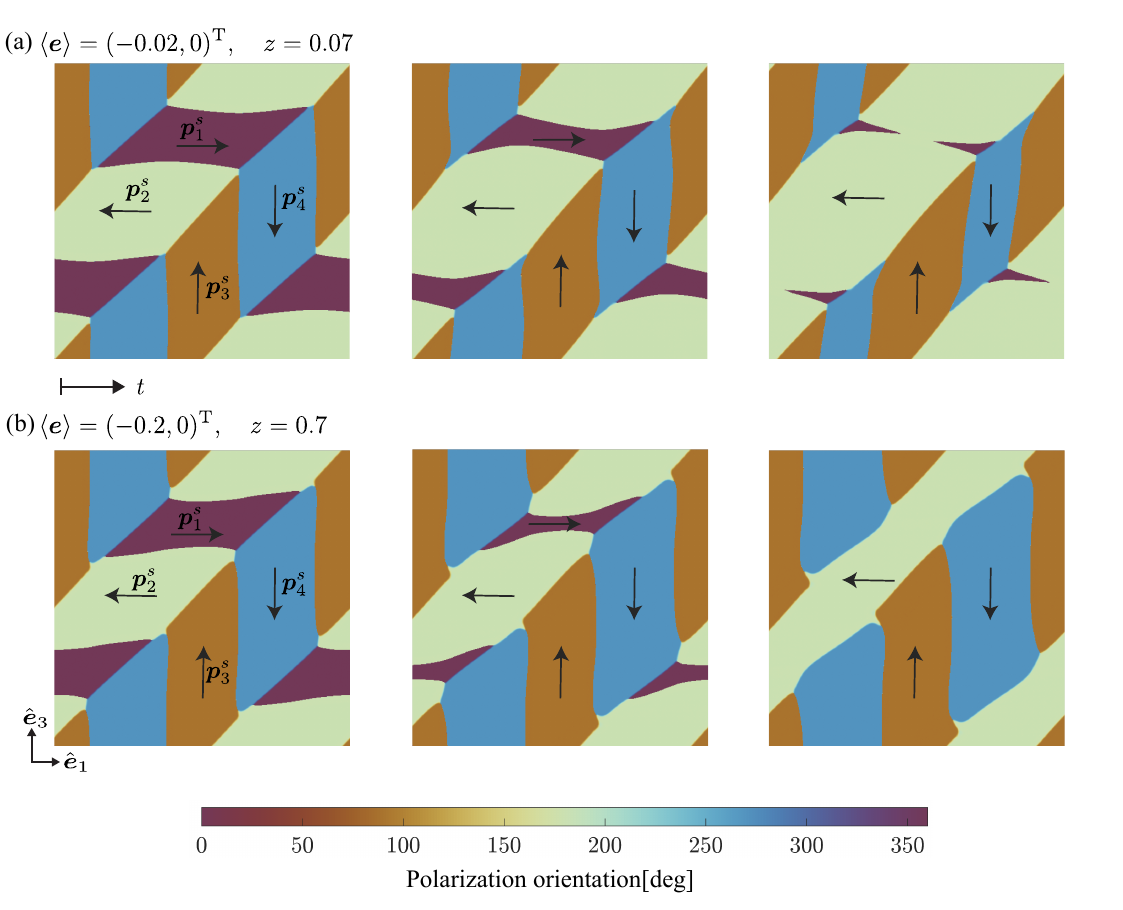}
    \caption{Disappearance of different domains depending on the magnitude of the applied electric field with progression of time from left to right. In (a), an electric field of magnitude 0.02 and zero average stress are applied to the rank-2 laminate domain structure favoring domains with $\bfp^s_2$. In (b), when the electric field magnitude is increased to 0.2, domains with $\bfp^s_3$ and $\bfp^s_4$ are favored, while domains with $\bfp^s_2$ are disfavored.} 
    \label{fig:anisoEffect}
\end{figure}

We illustrate this phenomenon in the fully coupled electromechanical BaTiO$_3$ setting using the general kinetic model with a linear kinetic relation for all domain walls. We again consider the rank-2 laminate microstructure of \autoref{fig:initPol} and the square RVE discretization from \autoref{tab:num_par}. We set $h=1$ and $g=0.5$, apply zero average stresses $\langle \bfsigma \rangle = \bfnull$, and apply different average electric fields.

In the first case\footnote{We focus on a qualitative comparison of domain evolution with two magnitudes of applied electric fields. We therefore choose different mobilities in the two cases: $\mu = 20$ for the low-field case and $\mu = 1$ for the high-field case. This choice reduces the total simulation time, because the driving force scales with the applied field, without a significant effect on the qualitative domain evolution.}, we prescribe $\langle \bfe \rangle = (-0.02,0)\T$ (corresponding to $3.7$~MV/m) and set $z=0.07$. \autoref{fig:anisoEffect}a shows that $\bfp^s_4$ is favored and dominates the evolving domain microstructure. In the second case, we prescribe $\langle \bfe \rangle = (-0.2,0)\T$ (37\,MV/m) and set $z=0.7$. For this case, \autoref{fig:anisoEffect}b shows the opposite trend, where $\bfp^s_3$ and $\bfp^s_4$ are favored. This indicates that, under sufficiently high applied electric fields and due to the strong anisotropy, the preferred domains can reverse, consistent with the transition predicted by the purely electrostatic model that led to~\eqref{eq:criticalValue}.

Note that we had to increase the magnitude of $z$ when applying the higher electric field to obtain stable numerical simulations. This highlights the importance of an appropriate choice of the parameter $z$ in the general kinetics model, which we further investigate in the following section.

\subsection{The effect of regularization using phase-fields} \label{sec:effect_regularization}

The presented multiphase-field formulation of the GKM is a numerically appealing alternative to the sharp-interface model for modeling ferroelectric domain evolution. The sharp-interface model can be implemented, e.g., via level-set methods \citep{cheng2025fft}, but the latter suffers from numerical complexity -- especially in the presence of multiple domain variants, which is exactly where the presented phase-field model offers a simple and efficient alternative. At the same time, the phase-field regularization introduces deviations from the sharp-interface model with effects on the simulation results. Therefore, in this section we examine the effect of the chosen phase-field regularization.

The dimensionless factor $z$, defined in \eqref{eq:reglarizedPar} and appearing in the regularized driving force \eqref{eq:drivingforce_dimless}, can be treated as the parameter that controls the degree of regularization imposed on the sharp-interface driving force. This, in turn, effects an increase in interface energy (as $z$ is directly propotional to $\Gamma$ in Eq. \eqref{eq:reglarizedPar}). Moreover, the weight on the gradient region increases. Therefore, $z$ not only controls the interface energy but also ensures a finite thickness of the regularized interface profile during its evolution, with consequences on the simulation outcome.

\begin{figure}[tb]
    \centering
    \includegraphics[width=1\linewidth]{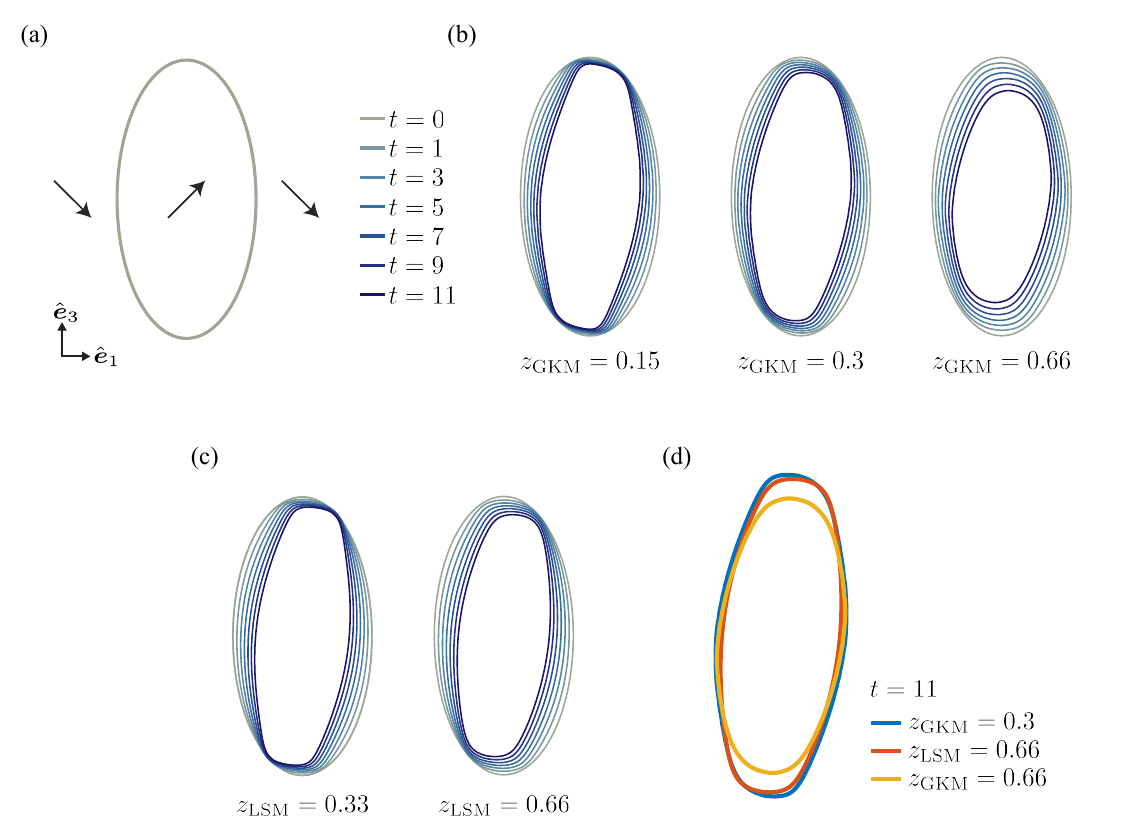}
    \caption{Effect of the regularization parameter $z$ on the evolution of a $90\degree$ domain nucleus in a square RVE with periodic boundary conditions. Zero average electric field and zero stress are applied to a nucleus with an initially elliptic shape (eccentricity 2), as shown in  (a). (b) Comparison of the nucleus evolution obtained from the phase-field model for $z=0.15$, $z=0.3$, and $z=0.66$ (at fixed $h=0.5$ and $g=1$). (c) Time evolution of the same nucleus simulated with the level-set method \citep{cheng2025fft}; the viscosity regularization is varied between $z_{\mathrm{LSM}} = 0.33$ and $z_{\mathrm{LSM}} = 0.66$. (d) Comparison of the nucleus shape at $t=11$ simulated by the general kinetic model ($z_{\mathrm{GKM}}$) and the level-set method ($z_{\mathrm{LSM}}$) for different regularization parameters.} 
    \label{fig:lsmgkm}
\end{figure}

\paragraph{Regularization introduces interface energy} To examine the effect on the interface energy, we simulate a 90$\degree$ domain nucleus of initially elliptic shape with an eccentricity of 2 in a periodic square RVE, as shown in \autoref{fig:lsmgkm}a. An average zero electric field and zero average stresses are applied, i.e., $\langle \bfe \rangle = \bfnull$ and $\langle \bfsigma \rangle = \bfnull$. This mimic the relaxation of a 90$\degree$ domain nucleus forming inside a uniform ferroelectric domain matrix. The parameters used to construct the RVE are included in \autoref{tab:num_par}. For simplicity, we prescribe a linear kinetic relation with unit mobility to the domain wall ($\check{V}_\beal(f_\beal) = f_\beal$). 
As shown in \autoref{fig:lsmgkm}b, increasing the regularization factor from $z=0.15$ to $z=0.66$ (while keeping all other parameters constant) increases the inward motion of the domain wall, driven by higher interface energy. We note that the gradient coefficient~$g$ is of minor importance here because multiphase junctions are absent. 

The comparison with the level-set method in \autoref{fig:lsmgkm}c and \autoref{fig:lsmgkm}d shows that the phase-field regularization has an effect similar to that in the level-set method, where adding interface energy to the sharp-interface driving force acts as a viscosity-like term in the evolution law that can ensure improved numerical stability. Moreover, the regularization parameter in the phase-field model and the nondimensionalized interface energy in the level-set method are of the same order of magnitude. Therefore, with an appropriate approximation, both methods yield qualitatively similar evolution, as shown in \autoref{fig:lsmgkm}d. For a thorough discussion of the connection between the evolution law of the level-set method and the phase-field model, we refer to \ref{apx:lsm_gkm}, which compares the two with a focus on nucleus evolution. 

\paragraph{Regularization controls the domain wall thickness} The regularized parameter $z$ in~\eqref{eq:reglarizedPar} (together with the material constants) also encodes the width of the regularized domain wall. In physical reality, the interface energy and the interface width are both material-dependent constants that can be determined, e.g., from first-principles simulations. In the phase-field context, both depend on the chosen discretization and model regularization, which is why one usually makes the choice to treat the energy as a physical and the width as a numerical parameter (see also the discussion in \citet{Guin2023phase} and \citet{feyen2023quantitative}). Consequently, for fixed interface energy, modifying $z$ can be regarded as changing the regularized interface width and hence the length scale of the problem. Here, setting $\Gamma$ to $0.1$~J/m$^2$ for barium titanate \citep{tagantsev2010, volker2011multiscale}, we obtain from \eqref{eq:reglarizedPar} the regularized thickness as $l = \Gamma/(z f_0 I) \approx (1.3\times10^{-9})/z$. This raises the question of the range of admissible values that parameter $z$ can assume to ensure a sufficient interface width for numerical stability.

To answer this question, recall that the purpose of the multi-obstacle potential in Eq.~\eqref{eq:ob_pot} is to maintain a finite regularized interface width during phase evolution\footnote{%
The purpose of the multi-obstacle potential is closely related to the reinitialization in the level-set method, where a reinitialization step is required to reconstruct the signed-distance function. This ensures the correct evaluation of the driving force.}. %
Therefore, the finite width fails when the bulk driving force is larger than the driving force contribution from the multi-obstacle potential. Equivalently, this implies that the bulk regions, in which the order parameter is supposed to be $0$ or $1$, strongly deviate from those desired minima \citep{garcke1999multiphaseConcept, feyen2023quantitative}. A lower bound on $z$ can hence be assessed by comparing the terms in the driving force \eqref{eq:drivingforce_dimless} of the two-phase interface, which leads to
\begin{equation}\label{eq:lowBound_reg}
    z > \lim_{\varphi \rightarrow 0} \frac{|f_b (\varphi)|}{ \Big| \frac{1}{h^2} \fp{\tilde\Phi(\varphi)}{\varphi}\Big| },
\end{equation}
where $f_b$ denotes the bulk driving force and $\tilde\Phi(\varphi)$ is the 2D slice of the multi-obstacle potential \eqref{eq:ob_pot} for the corresponding two-phase interface. The gradient term is absent, since it is zero in the bulk. 
\begin{figure}[!b]
    \centering
    \includegraphics[width=1\linewidth]{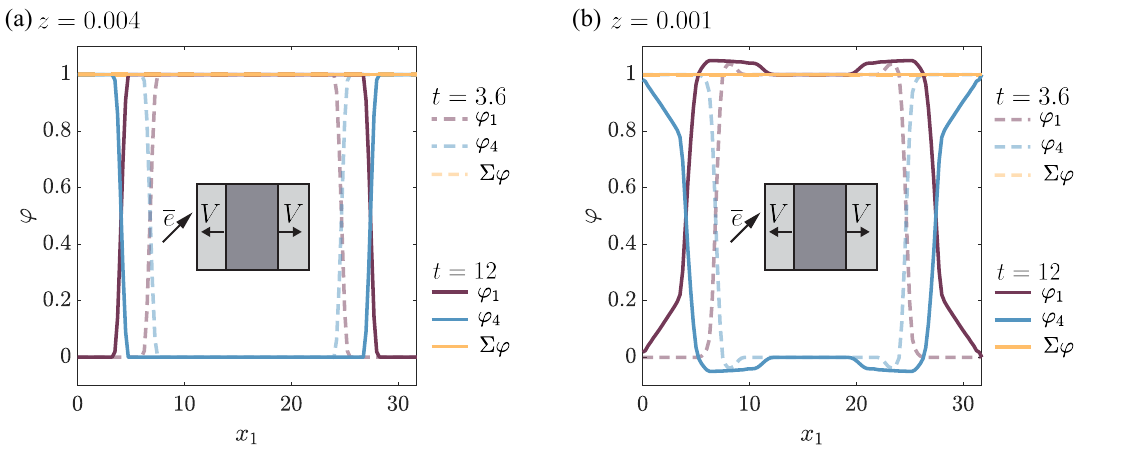}
    \caption{Demonstration of the necessary condition \eqref{eq:lowBound_reg} for maintaining a finite-thickness profile of order parameters in the phase-field model. An average electric field $\langle \bfe \rangle = \left(0.02/\sqrt{2}, 0.02/\sqrt{2}\right)\T$ and zero average stresses are applied to a straight, neutral $90\degree$ domain wall. The profiles of the order parameters $\{\varphi_1, \varphi_4\}$ and the sum of all order parameters $\sum_{\alpha=1}^4\varphi_\alpha$ are shown at two time instants (distinguished by opacity and solid vs.\ dashed). In (a), the profiles are well-maintained with $z=0.004$, whereas in (b) they fail with $z=0.001$. The lower bound in \eqref{eq:lowBound_reg} here evaluates to $0.003$.} 
    \label{fig:zlower_bound}
\end{figure}
The definition of the lower bound is analytically demonstrated in \ref{apx:lowerbound} with the simplified driving force that considers only isotropic electrostatic contributions. Here, we illustrate the numerical failure with the motion of a straight, neutral 90$\degree$ domain wall as in \autoref{fig:dw_180_90}b. We apply an average electric field $\langle \bfe \rangle = \left(0.02/\sqrt{2}, 0.02/\sqrt{2}\right)\T$, which is at a $45\degree$ angle to the positive $\hat{\bfe}_1$-axis, and zero average stresses $\langle \bfsigma \rangle = \bfnull$. These applied fields result in an initial (dimensionless) driving force of about $f_b=0.015$. The RVE is constructed with $N_1 \times N_3 = 64 \times 8$ grid points with $N_l=4$ in the regularized region. For the evolution law, we impose a linear kinetic relation with a mobility $\mu = 20$, and $h = 1$, $g = 0.5$ for the regularized driving force. Using \eqref{eq:lowBound_reg}, these parameters lead to the lower bound $z > 0.003$, with $\partial \tilde{\Phi}/\partial\varphi = 4 $, $f_b = 0.015$, and $h= 1$.  In  \autoref{fig:zlower_bound}a with $z = 0.004$, the interface propagates with a finite thickness of approximately $1$. In \autoref{fig:zlower_bound}b with $z = 0.001$, the finite thickness fails and the bulk regions creep away from the minima of $0$ and $1$. Consequently, the lower bound \eqref{eq:lowBound_reg} limits the interfacial thickness, which in turn affects the maximum physical size of the RVE that can be resolved. In this example, the lower bound of $z = 0.003$ corresponds to a maximum domain wall thickness of $400~ \mathrm{nm}$. Keeping the regularized parameter close to the lower bound generally allows the simulation of a large-scale problem with fewer grid points. We acknowledge that the lower bound on $z$ may further be reduced \citep{grose2022multi, feyen2023quantitative, zhang2023phase_voorhees}, which, however, goes beyond the scope of this study.

\section{Conclusion}
\label{sec:summary}
In the classical phase-field model of ferroelectrics (and beyond), the evolution law is typically given by an Allen--Cahn-type equation, a gradient descent of the energy. This evolution law implies a linear kinetic relation between the domain wall velocity and the driving force. In this work, we propose a multiphase-field model with a general kinetic formulation to study electromechanically coupled domain evolution in ferroelectrics. The proposed model enables an arbitrary (e.g., experimentally determined) kinetic relation to be incorporated directly into the evolution equation. We have demonstrated that the model can thus reproduce nonlinear kinetic relations of, e.g., mixed exponential--power law type as well as threshold-type kinetics. Moreover, within the multiphase-field framework, the evolution of a given order parameter is obtained as a combined interaction with the other order parameters. This pair interaction formulation enables not only separate kinetic relations to be assigned to different order parameter pairs, but also distinct interfacial energies to be prescribed for different interfaces. These capabilities are particularly useful for ferroelectrics, in which kinetic relations and interfacial energies are generally different among domain wall types. In addition, a stress-driven formulation is integrated in our multiphase-field framework, which allows the direct enforcement of average stresses and also utilizes the computational benefits of a stress-driven FFT solver. Overall, the presented general kinetic model provides a flexible framework for simulating and investigating the kinetics of interface motion in materials with applications beyond ferroelectrics (e.g., for solid-solid phase transformations). As with any model, the proposed general kinetic model can be further refined, for example by integrating an explicit nucleation scheme and by improving computational efficiency with a lower regularization requirement.

\section*{Acknowledgments}

The authors gratefully acknowledge the financial support from the Swiss National Science Foundation (SNSF) under project~212643. The authors appreciate the early investigations conducted by Laurent Guin and Louis Hennecart, and the discussion with Mathieu Brodmann, Claire Griesbach, Gerhard Br\"aunlich, Shashank Saxena, and Vignesh Kannan.   

\appendix


\section{Alternative representation of material tensors: example of the dielectric tensor}
\label{apx:reg_tensor}
In order to model spatial variations of the dielectric tensor, it is often constructed in the form of tensor products of $\bfp^s$, as shown, e.g., by \citet{schrade2013physical} and \citet{Guin2023phase}. In this section, we illustrate why this method is not adopted in our model. 
Using the spontaneous polarization as the basis vector such that $\hat{\bfe}_p = \bfp^s/p_0$, the dielectric tensor can be constructed as
\begin{align}
		\bfepsvar(\bfe_p) &= \epsilon_{c}(\hat{\bfe}_{p} \otimes \hat{\bfe}_{p} ) + \epsilon_{a}\big(\bfI - (\hat{\bfe}_{p}  \otimes \hat{\bfe}_{p} )\big) \nonumber\\%
		&= (\epsilon_{c} - \epsilon_{a}) (\hat{\bfe}_{p}\otimes \hat{\bfe}_{p}) + \epsilon_{a} \bfI.  \label{eq:form1_ep}
\end{align}
By expressing the polarization vectors through the phase-fields in \eqref{eq:pol_construct}, we can further rewrite \eqref{eq:form1_ep} as 
\begin{equation} \label{eq:epsvar}
	  \bfepsvar (\bfvarphi) = \frac{\epsilon_c - \epsilon_a}{p_0^2} \left( \sum_{\alpha=1}^4 \varphi_{\alpha} \bfp^s_{\alpha} \right) \otimes  \left(\sum_{\beta=1}^4 \varphi_{\beta}\bfp^s_{\beta} \right) + \epsilon_{a} \bfI.
\end{equation}
%
Now, consider a 180$^\circ$ domain wall that separates two domains, as shown in \autoref{fig:dw_180_90}a. For this two-phase interface, we write $\varphi_3$ = $\varphi$ and $\varphi_4 = 1-\varphi$, which simplifies \eqref{eq:epsvar} to yield the  throughout the domain as 
\begin{align}
    \bfepsvar (\varphi) &= (\epsilon_c - \epsilon_a) (2\varphi -1)^2 (\hat{\bfe}_3 \otimes \hat{\bfe}_3) + \epsilon_a \bfI. \label{eq:dieTensor_wrong}
\end{align}
In each domain (with $\varphi = 0$ or $\varphi = 1$), $\bfepsvar (\varphi)$ assumes correctly the identical values
\begin{equation} 
[\bfepsvar (1)] =  [\bfepsvar (0)] =
    \begin{pmatrix}
		\epsilon_{a}& 0 \\
		0& \epsilon_{c}
	\end{pmatrix}.
\end{equation}
However, in the middle of the regularized domain wall (where $\varphi=0.5$), the dielectric tensor becomes
\begin{equation}
[\bfepsvar (0.5)] = 
    \begin{pmatrix}
		\epsilon_{a}& 0 \\
		0& \epsilon_{a}
	\end{pmatrix},
\end{equation}
which implies a non-physical isotropic permittivity, in contrast to the correct constant dielectric tensor throughout the domains. This non-physical dielectric tensor causes erroneous domain evolution.

\section{Material parameters of single-crystalline BaTiO$_3$} \label{apx:Mat_BTO}
\setcounter{equation}{0} 
In this study, we choose barium titanate (BaTiO$_3$) as a representative ferroelectric material with a tetragonal crystal structure at room temperature. The material parameters of BaTiO$_3$ are listed in \autoref{tab:Mat_BTOPTO}, taken from \citet{devonshire1949_Part1}, \citet{li1991},  \citet{Jona1993Book}, and \citet{burcsu2001investigation_thesis}. As we use the stress-driven formulation in our simulations, the dielectric tensor at constant stress, $\bfEpsS$, is calculated from  $\bfEpsS = \bfepsvar + \sfbD \sfbC \sfbD\T$ \citep{IEEEStandard,cheng2025fft}, where  $\bfepsvar$ is the dielectric tensor at constant strain. In particular, the $c$-axis denotes the spontaneous polarization direction, and the $a$-axis denotes the other two directions that are orthogonal to the $c$-axis.  
\begin{table}[ht]
	\centering
	\small
	\setlength{\tabcolsep}{4pt}
	\caption{Material parameters for single-crystalline BaTiO$_3$ at room temperature ($25\degree\mathrm{C}$) under constant-stress conditions.}
	\label{tab:Mat_BTOPTO}
	\begin{tabular}{p{0.68\linewidth} >{\raggedleft\arraybackslash}p{0.28\linewidth}}
 	\hline
 	Spontaneous polarization ($\mathrm{C}/\mathrm{m^2}$) & \citep{Jona1993Book}\\
	         $p_0$ &   $0.26$  \\
  \hline
  Relative dielectric permittivity ($\epsilon_0 \approx 8.854 \times 10^{-12} ~\mathrm{V}/\mathrm{m}$) & \citep{li1991, cheng2025fft}\\
 	$\epsilon_c/\epsilon_0$ &  $157.66$ \\
 	$\epsilon_a/\epsilon_0$ &  $4115.22$\\
  \hline
  Elastic compliance moduli ($10^{-3}/\mathrm{GPa}$) & \citep{li1991}\\
 	$\sfS_{11}$ & $8.01$ \\
 	$\sfS_{33}$ &  $12.8$ \\
 	$\sfS_{13}$ &  $-4.6$ \\
 	$\sfS_{55}$ & $17.8$ \\
  \hline
  Piezoelectric strain constants ($10^{-12}\mathrm{C}/\mathrm{N}$)  & \citep{li1991} \\
 	$\sfD_{15}$  &  $580$ \\
 	$\sfD_{31}$  &  $-50$ \\
 	$\sfD_{33}$  & $106$  \\
  \hline
  Spontaneous strains & \citep{devonshire1949_Part1,burcsu2001investigation_thesis}\\
 	$\varepsilon_c$ &  $0.0067$ \\
    $\varepsilon_a$ &  $-0.0042$ \\
    \hline
\end{tabular}
\end{table}
The direction of the spontaneous polarization $\bfp^s$ determines the anisotropic directions of all material tensors \citep{Guin2023phase,cheng2024growth}. 

\section{Material tensors in their respective spontaneous polarization bases in the general kinetic model}\label{apx:tensorCons}
\setcounter{equation}{0}
In this section, we illustrate the dielectric, compliance, and piezoelectric tensors in their respective spontaneous polarization bases using Voigt notations. As we use the stress-driven formulation in our simulations, the dielectric tensor at constant stress, $\bfEpsS$, is calculated from  $\bfEpsS = \bfepsvar + \sfbD \sfbC \sfbD\T$ \citep{IEEEStandard,cheng2025fft}, where  $\bfepsvar$ is the dielectric tensor at constant strain. The dielectric permittivity along the $c$-axis is indicated by $\EpsS_{c}$, and the dielectric permittivity along the $a$-axis is indicated by $\EpsS_{a}$. The dielectric tensors $\bfEpsS_\alpha $ are given by
\begin{equation} \label{eq:dielec_phase}	[\bfEpsS_{1}]=[\bfEpsS_{2}] =
	\begin{pmatrix}
		\EpsS_{c}& 0 \\
		\mathrm{sym.} & \EpsS_{a}
	\end{pmatrix},  
\quad
	[\bfEpsS_{3}] = [\bfEpsS_{4}]=
\begin{pmatrix}
	\EpsS_{a}& 0 \\
	\mathrm{sym.} & \EpsS_{c}
\end{pmatrix},  
\end{equation}
where ``sym'' implies a symmetric matrix. The fourth-order compliance tensor $\sfbS$ has five independent constants in the three-dimensional case. In the two-dimensional setting, the only non-zero components are $\sfS_{11}$, $\sfS_{13}$, $\sfS_{33}$, and $\sfS_{55}$, where index 3 is the direction of the $c$-axis \citep{begun1949standards, JAFFE1971,li1991}. The compliance tensors $\sfbS_\alpha$ are hence given by
\begin{equation} \label{eq:compliance_phase}
	[\sfbS_{1}]=[\sfbS_{2}] =
	\begin{pmatrix}
		\sfS_{33} & \sfS_{13} & 0\\
		  & \sfS_{11} & 0 \\
		\mathrm{sym.} &  & \sfS_{55}
	\end{pmatrix},  
\quad
	[\sfbS_{3}] = [\sfbS_{4}]=
\begin{pmatrix}
		\sfS_{11} & \sfS_{13} & 0\\
		 & \sfS_{33} & 0 \\
		\mathrm{sym.} &  & \sfS_{55}
\end{pmatrix}.  
\end{equation}
The third-order strain-charge piezoelectric tensor $\sfbD$ has three independent constants, which are denoted by $\sfD_{15}$, $\sfD_{31}$, and $\sfD_{33}$. The piezoelectric tensors $\sfbD_\alpha$ are written as  
\begin{align} \label{eq:piezo_phase}
	&[\sfbD_{1}]=
	\begin{pmatrix}
		\sfD_{31} & \sfD_{33} & 0\\
		0 & 0 & \sfD_{15} \\
	\end{pmatrix},  
\quad 
	[\sfbD_{2}] = 
\begin{pmatrix}
		-\sfD_{31} & -\sfD_{33} & 0\\
		0 & 0 & -\sfD_{15} \\
	\end{pmatrix}, \\
	&[\sfbD_{3}]=
	\begin{pmatrix}
		0 & 0 & \sfD_{15} \\
		\sfD_{31} & \sfD_{33} & 0\\
	\end{pmatrix},  
\quad 
	[\sfbD_{4}] = 
\begin{pmatrix}
		0 & 0 & -\sfD_{15} \\
		-\sfD_{31} & -\sfD_{33} & 0\\
	\end{pmatrix}. 
\end{align}
We note that $\sfbD_\alpha$ of each phase is unique because of the electric field and the strain field coupling, as explained in \citet{kamlah2001}. Finally, the spontaneous strain is the strain arising from the crystalline structural change from the cubic phase to the tetragonal phase, in which the $c$-axis is stretched, while the other two axes are compressed by equal amounts. The spontaneous strains $\bfepsilon^s_\alpha$ in each phase take the form
\begin{equation} \label{eq:SS_phase}
	[\bfepsilon^s_{1}]=[\bfepsilon^s_{2}] =
	\begin{pmatrix}
		 \varepsilon_{c} & 0\\
        \mathrm{sym.} & \varepsilon_{a}
	\end{pmatrix}  ,
    \quad 
	[\bfepsilon^s_{3}]=[\bfepsilon^s_{4}] =
	\begin{pmatrix}
		 \varepsilon_{a} & 0\\
        \mathrm{sym.} & \varepsilon_{c}
	\end{pmatrix} .
\end{equation}

For a 180$\degree$ domain wall separating $\bfp^s=\pm p_0 \bfe_3$ as in \autoref{fig:dw_180_90}a, the spontaneous polarization and spontaneous strains in 2D in the two domains $\Omega_\alpha$ and $\Omega_\beta$ are given by
\begin{equation}
    [\bfp^s_\alpha] = \begin{pmatrix}
        0\\
        p_0
    \end{pmatrix}, \quad 
    [\bfp^s_\beta] = \begin{pmatrix}
        0\\
        -p_0
    \end{pmatrix}, \quad
    [\bfepsilon^s_\alpha] =[\bfepsilon^s_\beta] = \begin{pmatrix}
        \varepsilon_{a} & 0\\
        \mathrm{sym.} & \varepsilon_{c}
    \end{pmatrix}.
\end{equation}
The compliance tensors and dielectric tensors in the respective domains are written, using the Voigt matrix form \citep{nye1985physical}, as 
\begin{equation} 
[\sfbS_\alpha] = [\sfbS_\beta] =  
\begin{pmatrix}
	  \sfS_{11} &  \sfS_{13} & 0\\
		 & \sfS_{33} & 0\\
		\mathrm{sym.} &  & \sfS_{55}
  \end{pmatrix}, \quad [\bfEpsS_\alpha] = [\bfEpsS_\beta] =  
  \begin{pmatrix}
	  \EpsS_{a} &  0 \\
		\mathrm{sym.}  & \EpsS_{c}
  \end{pmatrix}.  
  \end{equation}
 The piezoelectric tensors in Voigt matrix form are given by 
\begin{equation}
  [\sfbD_\alpha]= 
  \begin{pmatrix}
            0 &  0 & \sfD_{15}\\
		 \sfD_{31}& \sfD_{33} & 0\\
  \end{pmatrix}, \quad [\sfbD_\beta] = 
  \begin{pmatrix}
		0 & 0 & -\sfD_{15} \\
		-\sfD_{31} & -\sfD_{33} & 0\\
	\end{pmatrix}.
\end{equation} 

For the vertical $90\degree$ domain wall in \autoref{fig:dw_180_90}b, whose crystal orientation is rotated by $45\degree$ from the horizontal/vertical axes, the spontaneous polarization and spontaneous strains in their respective domains $\Omega_\alpha$ and $\Omega_\beta$ (as defined in \autoref{fig:dw_180_90}) are 
\begin{equation}
    [\bfp^s_\alpha]= p_0
	\begin{pmatrix}
    \frac{1}{\sqrt{2}} \\[6pt]
	\frac{1}{\sqrt{2}}\\
	\end{pmatrix},  
    \quad
	[\bfp^s_\beta]= p_0
	\begin{pmatrix}
    \frac{1}{\sqrt{2}} \\[6pt]
	\frac{-1}{\sqrt{2}}\\
	\end{pmatrix}
\end{equation}
and
\begin{equation}
    [\bfepsilon^s_\alpha]=
	\begin{pmatrix}
\frac{\varepsilon_{a} +\varepsilon_{c}}{2} & \frac{\varepsilon_{c} -\varepsilon_{a}}{2}\\
	\mathrm{sym.} & \frac{\varepsilon_{a} +\varepsilon_{c}}{2}  \\
	\end{pmatrix},  
\quad
	[\bfepsilon^s_\beta]=
	\begin{pmatrix}
	\frac{\varepsilon_{a} +\varepsilon_{c}}{2} & \frac{\varepsilon_{a} -\varepsilon_{c}}{2}\\
	\mathrm{sym.} & \frac{\varepsilon_{a} +\varepsilon_{c}}{2}  \\
	\end{pmatrix}.
\end{equation}
The compliance tensors follow as 
\begin{align} \label{eq:elastic_phase_rot}
	&[\sfbS_{\alpha}]=
	\begin{pmatrix}
	  \frac{\sfS_{11}+2\sfS_{13}+\sfS_{33}+4\sfS_{55}}{4} &  \frac{\sfS_{11}+2\sfS_{13}+\sfS_{33}-4\sfS_{55}}{4} & \frac{-\sfS_{11}+\sfS_{33}}{4}\\[6pt]
		 & \frac{\sfS_{11}+2\sfS_{13}+\sfS_{33}+4\sfS_{55}}{4} & \frac{-\sfS_{11}+\sfS_{33}}{4}\\[6pt]
		\mathrm{sym.} &  & \frac{\sfS_{11}-2\sfS_{13}+\sfS_{33}}{4}
	\end{pmatrix},  \nonumber\\
	 &[\sfbS_\beta]=
\begin{pmatrix}
	 \frac{\sfS_{11}+2\sfS_{13}+\sfS_{33}+4\sfS_{55}}{4} &  \frac{\sfS_{11}+2\sfS_{13}+\sfS_{33}-4\sfS_{55}}{4} & \frac{\sfS_{11}+\sfS_{33}}{4}\\[6pt]
		 & \frac{\sfS_{11}+2\sfS_{13}+\sfS_{33}+4\sfS_{55}}{4} & \frac{\sfS_{11}+\sfS_{33}}{4}\\[6pt]
		\mathrm{sym.} &  & \frac{\sfS_{11}-2\sfS_{13}+\sfS_{33}}{4} \nonumber
\end{pmatrix}. 
\end{align}
The dielectric tensors have components
\begin{align}
	[\bfEpsS_\alpha]=
	\begin{pmatrix}
		\frac{\EpsS_{a}+\EpsS_{c}}{2} & \frac{-\EpsS_{a}+\EpsS_{c}}{2} \\[6pt]
		\mathrm{sym.}& \frac{\EpsS_{a}+\EpsS_{c}}{2} 
	\end{pmatrix},  
\quad
	[\bfEpsS_\beta]=
	\begin{pmatrix}
		\frac{\EpsS_{a}+\EpsS_{c}}{2} & \frac{\EpsS_{a}-\EpsS_{c}}{2} \\[4pt]
		\mathrm{sym.}& \frac{\EpsS_{a}+\EpsS_{c}}{2} 
	\end{pmatrix}.    \nonumber 
\end{align}
The piezoelectric tensors are
\begin{align}
	&[\sfbD_\alpha]=
	\begin{pmatrix}
		\frac{2\sfD_{15}+\sfD_{31}+\sfD_{33}}{2\sqrt{2}} & \frac{-2\sfD_{15}+\sfD_{31}+\sfD_{33}}{2\sqrt{2}}  &     \frac{-\sfD_{31}+\sfD_{33}}{2\sqrt{2}} \\[6pt]
		\frac{-2\sfD_{15}+\sfD_{31}+\sfD_{33}}{2\sqrt{2}} & \frac{2\sfD_{15}+\sfD_{31}+\sfD_{33}}{2\sqrt{2}} & \frac{-\sfD_{31}+\sfD_{33}}{2\sqrt{2}}\\
	\end{pmatrix},   \nonumber\\[6pt]
	&[\sfbD_\beta]=
	\begin{pmatrix}
		\frac{2\sfD_{15}+\sfD_{31}+\sfD_{33}}{2\sqrt{2}} & \frac{-2\sfD_{15}+\sfD_{31}+\sfD_{33}}{2\sqrt{2}} & \frac{\sfD_{31}-\sfD_{33}}{2\sqrt{2}}\\[6pt]
		\frac{2\sfD_{15}-\sfD_{31}-\sfD_{33}}{2\sqrt{2}} & \frac{-2\sfD_{15}-\sfD_{31}-\sfD_{33}}{2\sqrt{2}} & \frac{-\sfD_{31}+\sfD_{33}}{2\sqrt{2}} 
	\end{pmatrix}.   \nonumber 
\end{align}
\section{Stability of the general kinetic model: specialization to isotropic electrostatics}\label{apx:lowerbound}
\setcounter{equation}{0} 
\begin{figure}[!b]
    \centering
    \includegraphics[width=1\linewidth]{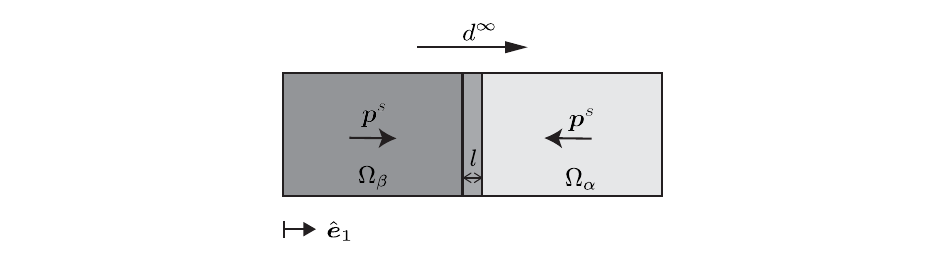}
    \caption{Schematic of a charged 180$\degree$ domain wall with a domain wall thickness $l$ in a 1D configuration. The spontaneous polarization changes from $p_0\hat{\bfe}_1$ to $- p_0 \hat{\bfe}_1$ across the domain wall. A constant electric displacement $d^\infty$ is applied remotely. } 
    \label{fig:180Charged}
\end{figure}
The lower bound specified in \eqref{eq:lowBound_reg} results from the fact that the multi-obstacle potential ensures the finite thickness of the domain wall, enforcing the values of $0$ and $1$ in the bulk. The exact value of the lower bound can be determined analytically in the simplified case of considering only isotropic electrostatics in the evolution of an electrically charged $180\degree$ domain wall in 1D. This 1D configuration exhibits a spontaneous polarization that varies from $p_0\hat{\bfe}_1$ to $- p_0 \hat{\bfe}_1$ across the domain wall (where $p_0$ is normalized to $1$ by the spontaneous polarization magnitude of the given material). A schematic of this scenario is shown in \autoref{fig:180Charged}. Since only two phase variants are involved, we let $\varphi_\alpha = \varphi$ and $\varphi_\beta = 1-\varphi$. This gives the dimensionless regularized driving force written as 
\begin{equation}\label{eq:reg_df_1d}
    f = f_b  + z \left( - \frac{1}{h^2} \fp{\tilde\Phi(\varphi)}{\varphi} + 2\nabla^2\varphi\right),
\end{equation}
where $f_b = \calG_\beta - \calG_\alpha$ denotes the bulk driving force in \eqref{eq:drivingforce_ferro}. In the isotropic electrostatic case, the bulk energy contribution $f_b$ simplifies to $f_b =\bfe \cdot (\bfp^s_\beta - \bfp^s_\alpha) = -2e$. Let us assume that a constant (dimensionless) electric displacement $d^\infty$ is applied remotely. The electric field is thus obtained as \citep{Guin2023phase}
\begin{equation}
    e = d^\infty - p = d^\infty - (1-2\varphi),
\end{equation}
where $p = \varphi p^s_\alpha + (1-\varphi) p^s_\beta = -\varphi + (1-\varphi)$, since $\bfp^s_\alpha = -1$ and $\bfp^s_\beta = 1$. Substituting the obtained $f_b$ into the driving force \eqref{eq:reg_df_1d} leads to 
\begin{equation}
    f = -2\left[d^\infty - (1-2\varphi)\right] + z\left( - \frac{1}{h^2} \fp{\tilde\Phi(\varphi)}{\varphi} + 2\nabla^2\varphi\right).
\end{equation}
At the region where the domain wall connects to the bulk ($\varphi = 0$), the regularized driving evaluates to
\begin{equation}\begin{aligned}
         \lim_{\varphi\rightarrow 0 } f &= \lim_{\varphi\rightarrow 0 } \big[-2\big(d^\infty  - (1-2\varphi)\big) + z\bigg(- \frac{4(1-2\varphi)}{h^2}\bigg) + 2\nabla^2\varphi \big],\\
        &= -2(d^\infty -1) + z (-4).
\end{aligned}\label{eq:lowerBoundPhi0}
\end{equation}
where $h=1$ and $\nabla^2 \varphi = 0$ are assumed. 

Noticing $\pl\tilde\Phi(\varphi)/\pl\varphi = 4 (1-2\varphi)$, the regularized driving force can be written as 
\begin{equation} \label{eq:effectivePotential}
\begin{aligned}
        f &= -2 d^\infty + \frac{1}{2} \fp{\tilde\Phi(\varphi)}{\varphi} + z\left( - \frac{1}{h^2} \fp{\tilde\Phi(\varphi)}{\varphi} + 2\nabla^2\varphi\right), \\
         &= -2 d^\infty + z\left( \left(- \frac{1}{h^2} + \frac{1}{2z}\right)  \fp{\tilde\Phi(\varphi)}{\varphi} + 2\nabla^2\varphi\right).
\end{aligned}
\end{equation}
To ensure the effectiveness of the double-obstacle potential even when $d^\infty = 0$, it requires that 
\begin{equation}
    \left(- \frac{1}{h^2} + \frac{1}{2z}\right)  > 0.  
\end{equation}
For $h=1$, we arrive at the lower bound
\begin{equation} 
    z > 0.5,
\end{equation}
which ensures the existence of the double-obstacle potential.
Together with \eqref{eq:lowerBoundPhi0}, this shows that the lower bound on $z$, required for the stability of the scheme, is given by
\begin{equation} \label{eq:zBound_1D}
    z > \frac{\lvert d^\infty - 1\rvert}{2} > 0.5,
\end{equation}
which must be respected even when $d^\infty = 0$. The lower bound on $z$ for a general bulk driving force (going beyond electrostatics) is given by
\begin{equation}
     z > \lim_{\varphi \rightarrow 0} \frac{|f_b (\varphi)|}{ \Big| \frac{1}{h^2} \fp{\tilde\Phi(\varphi)}{\varphi}\Big| }.
\end{equation}
%

\section{The connection between the evolution law in the level-set method and the general kinetic model}\label{apx:lsm_gkm}
\setcounter{equation}{0} 
In general, the evolution law \eqref{eq:evo_phi} in our general kinetic model (GKM) is derived as the propagation of level sets \citep{alber2016alternative}. This derivation leads to a form that closely resembles the evolution law in the level-set method (LSM) \citep{sethian1996fast,osher2004level}, as discussed in \autoref{sec:multiphase}. The connection between these two distinct evolution laws can be cleanly analyzed for the special case of two-phase evolution. In particular, we assume a linear kinetic relation with unit mobility for the discussion in this section.  In the general kinetic model, the evolution law for the two-phase interface in the dimensionless form is written as 
\begin{equation} \label{eq:gkm2phase}
    \parT \varphi = \lvert \nabla \varphi \rvert \bigg[f_b  + z_{\mathrm{GKM}}\left( - \frac{1}{h^2} \fp{\tilde\Phi(\varphi)}{\varphi} + 2\nabla^2\varphi\right)\bigg],
\end{equation}
where $f_b$ is the bulk driving force as defined in \eqref{eq:drivingforce_ferro}, $\tilde{\Phi} (\varphi)$ the 2D slice of the multi-obstacle potential, and $z_{\mathrm{GKM}}$ the dimensionless parameter associated with the general kinetic model. Here, we choose $g = 1$ for simplicity.

The level-set evolution law with viscosity through interface energy \citep{zhao1996variational_Osher} takes the form 
\begin{equation}\label{eq:levelset_evo0}
    \parT\psi =  f_b \lvert \nabla \psi\rvert + \Gamma \nabla^2\psi, 
\end{equation}
where $f_b$ is the same bulk driving force as defined in \eqref{eq:drivingforce_ferro}\footnote{%
We note that the driving force convention in the general kinetic model is defined opposite to that of the level-set method. For convenience, we here retain our general kinetic model convention to formulate \eqref{eq:levelset_evo0}.   
}.~%
$\psi$ is the signed-distance function, so $\Gamma \nabla^2\psi$ is the product of the interface energy and the local interface curvature. We first nondimensionalize \eqref{eq:levelset_evo0} as follows:
\begin{equation}
    \ol{f}_b = \frac{f_b}{f_0}, \quad \ol{x} = \frac{x}{ l_0}, 
    \quad \ol{\psi} = \frac{\psi}{l_0}, \quad \ol{t} = \frac{t}{1/f_0},
\end{equation}
with $l_0$ being the thickness of the interface, and $f_0 = p_0^2/\EpsS_{c}$ the characteristic driving force defined in \eqref{eq:reglarizedPar}. Nondimensionalization in this particular case is simplified with the assumption of unit mobility. With these definitions, we arrive at the dimensionless evolution equation of the level-set method: 
\begin{equation}\label{eq:levelset_evo}
    \parT\psi = f_b \lvert \nabla \psi\rvert + z_{\mathrm{LSM}} \nabla^2\psi,
\end{equation}
where we dropped the overhead bar for conciseness, and we defined the regularization parameter of the level-set method as 
\begin{equation}
    z_{\mathrm{LSM}} = \frac{\Gamma}{f_0 l_0}. 
\end{equation}

In the following, we directly work with dimensionless quantities. Using the equilibrium profile of the phase-field in \eqref{eq:equiProfile}, we perform a change of variables \citep{glasner2001nonlinear, sun2007sharp}, so that
\begin{equation}
    \psi = \frac{h}{2}\sin\me(2\check\varphi - 1),
\end{equation}
where $h$ is the ratio between the characteristic length $l_0$ and the thickness $l$ of the interface in the phase-field (see Eq.~\eqref{eq:nonDimension_evoLaw}). Taking the gradient leads to the relation 
\begin{equation}
    \lvert \nabla \psi \rvert = \frac{h}{2} \frac{1}{\sqrt{\varphi (1-\varphi)}}\lvert \nabla \check\varphi \rvert.  
\end{equation}
Since $\psi$ is a signed-distance function, we know that $\lvert\nabla \psi\rvert = 1$, which simplifies the above to  
\begin{equation}\label{eq:lsmgkm_relation}
    \lvert \nabla \check\varphi \rvert = \frac{2}{h} \sqrt{\check\varphi (1-\check\varphi)}, \quad \parT \psi = \frac{1}{\lvert \nabla \check\varphi\rvert} \parT \check\varphi, \quad \nabla^2 \psi = -\frac{2}{h^2}(1-2\check\varphi) + \nabla^2\check\varphi.
\end{equation}
Notice that we can further simplify the last term by acknowledging that $4 (1-2\varphi)=\tilde\Phi'(\varphi)$ is the derivative of the 2D slice of the multi-obstacle potential for the present two-phase scenario. Making this substitution and further inserting the relations \eqref{eq:lsmgkm_relation} into the evolution law of the level-set \eqref{eq:levelset_evo} lets us rewrite the latter as
\begin{equation}\label{eq:LSMToPFM}
    \parT \check\varphi = f_b \lvert \nabla \check\varphi\rvert  + z_{\mathrm{LSM}}\left(- \frac{1}{2h^2} \fp{\tilde\Phi(\varphi)}{\varphi} + \nabla^2\check\varphi\right).
\end{equation}
In Eq.~\eqref{eq:LSMToPFM}, the emergence of the double-obstacle potential stems from the assumption that the level-set function remains a signed-distance function at all times. This helps explain why phase-field models require a nonconvex potential, whereas level-set methods require regular reinitialization.

Eq.~\eqref{eq:LSMToPFM} is similar to the analogous relation in the general kinetic model, given by Eq.~\eqref{eq:gkm2phase}, except that the regularized term is missing the prefactor $\lvert \nabla \check \varphi \rvert$. 
Despite this difference, we note that $\lvert \nabla \check \varphi \rvert$ takes a value between 0 and 1, which means the evolution close to the center of the regularized interface is similar to the general kinetic model. This allows us to address the connection between the dimensionless parameters $z_{\mathrm{LSM}}$ and $z_{\mathrm{GKM}}$ in the two models. When we choose the characteristic length such that $h=1/\sqrt{2}$, the evolution law in \eqref{eq:LSMToPFM} becomes 
\begin{equation}\label{eq:LSMToPFM_h}
    \parT \check\varphi = f_b \lvert \nabla \check\varphi\rvert  + z_{\mathrm{LSM}}\left(-\fp{\tilde\Phi(\check\varphi)}{\check\varphi} + \nabla^2\check\varphi\right), 
\end{equation}
which leaves only $z_{\mathrm{LSM}}$ as the free parameter. 


Comparing the regularized parameter $z_{\mathrm{LSM}}$ to $z_{\mathrm{GKM}}$ from \eqref{eq:reglarizedPar}, we notice that 
\begin{equation}\label{eq:relationZ}
    z_{\mathrm{GKM}} = \frac{\Gamma h^2}{f_0 l I} = \frac{\Gamma h }{ f_0 l_0 I} =  z_{\mathrm{LSM}} \frac{h}{I}
    \quad\Rightarrow\quad z_{\mathrm{GKM}} = \frac{\sqrt{2}}{\pi} z_{\mathrm{LSM}} \approx 2.22\, z_{\mathrm{LSM}},
\end{equation}
where we exploited that $I = \pi/2$ and $h$ is chosen to be $1/\sqrt{2}$. 
This relation motivates us to compare the nucleus evolution obtained from the level-set method and the general kinetic model -- with the expectation that, if \eqref{eq:relationZ} is satisfied, the two models yield qualitatively similar behavior. For comparison, \autoref{fig:lsmgkm} shows simulation results from both the general kinetic phase-field model and the analogous level-set formulation \citep{cheng2025fft}. The direct comparison in \autoref{fig:lsmgkm}d illustrates that the two models indeed produce similar results when comparing $z_{\mathrm{LSM}} = 0.66$ and $z_{\mathrm{GKM}} = 0.33$, which approximately satisfy \eqref{eq:relationZ}. By contrast, results for $z_{\mathrm{LSM}} = z_{\mathrm{GKM}} = 0.66$ differ significantly.

In summary, we have illustrated the connection between the impact of regularization in the general kinetic (phase-field) model and the effect of viscosity in the level-set method. It is important to note that the comparison provided herein is a qualitative approximation intended to illustrate the similarity between the two methods. This is especially the case when the viscosity term is required in the level-set method to simulate multiphase evolution \citep{zhao1996variational_Osher}.

\newpage
 \bibliographystyle{elsarticle-harv} 
 \bibliography{HC_phD}





\end{document}